%% file: arxiv_frames.tex
\documentclass[a4paper,12pt]{article}

\usepackage{amsmath,amssymb,graphicx,amsthm,ifthen,epstopdf,fancyhdr,color}
\usepackage[margin=3cm,vmargin={2cm,3.5cm},includefoot]{geometry}
\usepackage[bookmarks=true]{hyperref}
\usepackage{booktabs}
\usepackage{palatino,rotating}
\usepackage{cite}
\usepackage{fancyhdr}
\DeclareMathOperator{\tr}{tr}

\newcommand\x{\times}
\newcommand{\V}[1]{\ensuremath{\mathbf{#1}}}

\newcommand{\R}{\ensuremath{\mathbb{R}}}
\newcommand{\E}{\ensuremath{\mathbb{E}}}
\newcommand{\C}{\ensuremath{\mathbb{C}}}
\newcommand{\Fc}{\ensuremath{\mathcal{F}}}

\newcommand{\m}{m}

\newcommand{\norm}[1]{\left|\left| #1 \right|\right|}

\newcommand{\specstat}{\ensuremath{\Psi}}
\newcommand{\vx}{\ensuremath{\mathbf{x}}}
\newcommand{\Xk}{\ensuremath{X_K}}
\newcommand{\Xkn}{\ensuremath{X_{K_n}}}
\newcommand{\Gk}{\ensuremath{G_K}}
\newcommand{\Gkn}{\ensuremath{G_{K_n}}}

\begin{document}

\title{Random Subsets of Structured Deterministic Frames have MANOVA Spectra}

\author{
    Marina Haikin \footnotemark[1]
    \and 
    Ram Zamir \footnotemark[1]
    \and
    Matan Gavish \footnotemark[2]
}

\date{}
\maketitle

\renewcommand{\thefootnote}{\fnsymbol{footnote}}
\footnotetext[1]{EE - Systems Department, Tel Aviv University, Tel Aviv, Israel}
\footnotetext[2]{School of Computer Science and Engineering, Hebrew University of
Jerusalem}
\renewcommand{\thefootnote}{\arabic{footnote}}

\begin{abstract}
We draw a random subset of $k$ rows from a frame with $n$ rows (vectors) and $m$
columns (dimensions), where $k$ and $m$ are proportional to $n$.  For a variety
of important deterministic equiangular tight frames (ETFs) and tight non-ETF
frames, we consider the distribution of singular values of the $k$-subset
matrix.  We observe that for large $n$ they can be precisely described by a
known probability distribution -- Wachter's MANOVA spectral distribution, a
phenomenon that was previously known only for two types of random frames.  In
terms of convergence to this limit, the $k$-subset matrix from all these frames
is shown to be empirically indistinguishable from the classical MANOVA (Jacobi) random matrix
ensemble.  
Thus empirically the MANOVA ensemble offers a universal description
of the spectra of randomly selected
$k$-subframes, even those taken from deterministic frames. The same
universality phenomena is
shown to hold for notable random frames as well.
This description enables exact calculations of properties of solutions for
systems of linear equations based on a random choice of $k$ frame vectors out of
$n$ possible vectors, and has a variety of implications for erasure coding,
compressed sensing, and sparse recovery.  
When the aspect ratio $m/n$ is small, the MANOVA spectrum tends to the well
known Mar\u cenko-Pastur distribution of the singular values of a Gaussian
matrix, in agreement with previous work on highly redundant frames.
Our results are empirical, but they
are exhaustive, precise and fully reproducible.
\end{abstract}


\noindent
Consider a frame $\{\vx_i\}_{i=1}^n \subset \R^\m$ or $\C^\m$ 
and stack the vectors as rows to obtain the $n$-by-$\m$ frame matrix $X$. 
Assume that $\norm{
\vx_i}_2=1$ (deterministic frames) or 
$\lim_{n\to \infty}\|\vx_i\|=1$ almost surely (random frames). 
This paper studies properties of a random subframe 
$\{\vx_i\}_{i\in K}$, where $K$ is chosen uniformly at random 
from $[n]=\left\{ 1,\ldots,n \right\}$ and $|K|=k\leq n$.
We let $\Xk$ denote  
the $k$-by-$\m$ submatrix of $X$ created by picking only
the rows $\{\vx_i\}_{i\in K}$; call this object {\em a typical $k$-submatrix
of $X$}.
We consider 
a collection of well-known 
deterministic frames, listed in Table \ref{frames:tab}, 
which we denote by $\cal X$. Most of the frames in $\cal
X$ are equiangular tight frames (ETFs), and some are near-ETFs.

This paper suggests that for a frame in $\cal X$ it is possible to calculate
quantities of the form $\E_K \specstat(\lambda(\Gk))$, 
where $\lambda(\Gk)=(\lambda_1(G_K),...,\lambda_k(G_K))$ is the
vector of eigenvalues of the $k$-by-$k$ Gram matrix $\Gk=\Xk \Xk'$
and $\specstat$ is a functional of these eigenvalues.
As discussed below, such quantities are of considerable interest in various
applications where frames are used, across a variety of domains, including
compressed sensing, sparse
recovery and erasure coding.

We present a simple and explicit formula for calculating 
$\E_K \specstat(\lambda(\Gk))$ for a given frame in $\cal X$ and a given spectral 
functional $\specstat$. Specifically, for the case $k \le m$,
\[
\E_K\specstat(\lambda(\Gk)) \approx \specstat\left( f^{MANOVA}_{\beta,\gamma}
\right)\,, \] 
where $\beta=k/\m$, $\gamma=\m/n$ and where 
$f^{MANOVA}_{\beta,\gamma}$ is the density of
Wachter's classical 
MANOVA$(\beta,\gamma)$
limiting distribution \cite{wachter}. The fluctuations about this approximate 
value are given {\em exactly} by
\begin{eqnarray} \label{main_for:eq}
\E_K \big| \specstat(\lambda(\Gk)) - \specstat\left( f^{MANOVA}_{\beta,\gamma}
\right)\big|^2 =   
C n^{-b} \log^{-a}(n) \,.
\end{eqnarray}
While the constant $C$ may depend on the frame, 
the exponents $a$ and $b$ are {\em universal} and depend only 
on $\specstat$ and on the aspect ratios $\beta$ and $\gamma$. 
Evidently, the precision of the MANOVA-based approximation is good, known, and 
improves as $m$ and $k$ both grow proportionally to $n$.

Formula \ref{main_for:eq} is based on a far-reaching 
{\em universality hypothesis}: 
For all frames in $\cal X$, as well as for well-known random frames also listed in
Table \ref{frames:tab}, we find that
the spectrum of the typical $k$-submatrix ensemble is indistinguishable 
from that of the classical MANOVA (Jacobi) random matrix ensemble
\cite{forrester}
of the same
size. 
(Interestingly, it will be shown that for deterministic ETFs this
indistinguishably holds in a stronger sense than for deterministic non-ETF
frames.)
This universality is not asymptotic, and concerns 
finite $n$-by-$\m$ frames. However, it does imply that the spectrum of the
typical $k$-submatrix ensemble converges to 
a  universal limiting distribution, which is non other than 
Wachter's  
MANOVA$(\beta,\gamma)$ limiting distribution \cite{wachter}. 
It also implies that the universal exponents $a$ and $b$
in \eqref{main_for:eq}
are previously unknown, universal quantities corresponding to the 
classical MANOVA (Jacobi) random matrix ensemble.


This brief announcement tests Formula
\ref{main_for:eq} and the underlying universality hypothesis
by conducting substantial computer
experiments, in which a large number of random $k$-submatrices are generated. 
We  study a large variety of  deterministic frames, both
real and complex. In addition to the universal object (the MANOVA ensemble)
itself, 
we study difference-set spectrum frames,
Grassmannian frames, real Paley frames, complex Paley frames, quadratic phase
chirp frames, Spikes and Sines frames, and Spikes and Hadamard frames. 

We report
compelling empirical evidence,  systematically documented and 
analyzed,
which fully supports the universality hypothesis and \eqref{main_for:eq}.
Our results are empirical, but they are exhaustive, 
precise, reproducible and meet the best standards 
of empirical science. 

For this purpose, we develop a natural
framework for empirically testing such hypotheses regarding
limiting distribution and convergence rates of 
random matrix ensembles. 
Before turning to deterministic frames, 
we validate our framework on well-known random frames, including 
real orthogonal Haar frames, complex unitary Haar frames, real random
Cosine frames and complex random Fourier frames.
Interestingly, rigorous proofs that identify the MANOVA distribution as the
limiting spectral distribution of typical $k$-submatrices can be found in the
literature for two of these random frames, namely the random Fourier frame
\cite{Farrell} and the unitary Haar frame \cite{Edelman}. 
\section*{Motivation} \label{sec:motivation} 

Frames can be viewed as an analog counterpart for digital coding. They provide 
overcomplete representation of signals, adding redundancy and increasing
immunity to noise. 
Indeed, they are used in
many branches of science and engineering for stable signal representation,
as well as error and erasure correction.

Let $\lambda(G)$ denote the vector of nonzero eigenvalues of $G=X'X$ and let 
$\lambda_{max}(G)$ and $\lambda_{min}(G)$ denote its max and min,
respectively. 
Frames were traditionally designed to achieve frame bounds  $\lambda_{min}(G)$
as high as possible (resp. $\lambda_{max}(G)$ as low as possible).
Alternatively, they were designed to
minimize  {\em mutual coherence} \cite{Donoho-stable,Elad2010},
the maximal pairwise correlation between any two frame vectors.

In the passing decade it has become
apparent that neither frame bounds (a global criterion) 
nor coherence (a local, pairwise criterion) are
sufficient to
explain various phenomena related to overcomplete representations, and that
one should also look at collective behavior of $k$ frame vectors from the
frame, $2\leq k \leq n$. 
%
%
%
While different applications focus on different properties of the submatrix
$\Gk$, most of these properties can be expressed as a function of
$\lambda(\Gk)$,
and even just an average of a scalar function of the eigenvalues. Here are a few notable examples. 
\paragraph{Restricted Isometry Property (RIP).} 
Recovery of any $k/2$-sparse
signal $\V{v}\in\R^n$ from its linear measurement $F'\V{v}$ using $\ell_1$
minimization is guaranteed if the spectral radius of $\Gk - I$, namely,
\begin{eqnarray} \label{rip_func:eq}
\specstat_{RIP}(\lambda(\Gk))= 
\max\{ \lambda_{max}(\Gk)-1 \,,\,
1-\lambda_{min}(\Gk) \}\,, 
\end{eqnarray} 
is uniformly bounded by some $\delta<0.4531$ on all $K\subset[n]$
\cite{CT06, Candes-RIP,Lai}.

\paragraph{Statistical RIP.} Numerous authors have studied a relaxation of the
RIP condition suggested in \cite{Caulderbank-STRIP}. Define 
\begin{eqnarray} \label{strip_func:eq}
\specstat_{StRIP,\delta}(\lambda(\Gk)) =
\begin{cases} 1 &  \specstat_{RIP}(\lambda(\Gk)) \leq
\delta \\ 
0 & otherwise \end{cases}\,.  
\end{eqnarray}
Then
$\E_K \specstat_{StRIP,\delta}(\lambda(\Gk))$ is
the probability that the RIP condition 
with bound $\delta$ holds when $X$ acts on a signal
supported on a random set of $k$ coordinates.
%
%
\paragraph{Analog coding of a source with erasures.} 
In \cite{AnalogCoding} two of us  
considered a typical erasure pattern of
$n-k$ random samples known at the transmitter, but not the receiver.
The rate-distortion function  
of the coding scheme suggested in \cite{AnalogCoding} is determined
by $\E_K \log(\beta\specstat_{AC}(\lambda(\Gk)))$, with
%
\begin{eqnarray} \label{ac_func:eq}
\specstat_{AC}(\lambda(\Gk))=\frac{1}{k}\tr[(\Gk)^{-1}]/\left(\frac{1}{k}\tr[\Gk]\right)^{-1}\,,
\end{eqnarray}
i.e.,$\specstat_{AC}(\lambda(\Gk))$ is the arithmetric-to-harmonic means ratio of the eigenvalues (the arithmetric mean is $1$ due to the normalization of frames). This quantity is the signal
amplification responsible for the excess rate of the suggested coding scheme. Note that $\beta$ here is the inverse of $\beta$ defined in \cite{AnalogCoding}.  
\paragraph{Shannon transform.} 
The quantity 
\begin{eqnarray} \label{shannon_func:eq}
\begin{aligned}
\specstat_{Shannon}(\lambda(\Gk)) 
&= \frac{1}{k}\log(\det(I+\alpha \Gk))\\
&= \frac{1}{k}\tr(\log(I+\alpha \Gk))\,,
\end{aligned}
\end{eqnarray}
which was suggested in \cite{RandomMatrix},
measures the capacity of a linear-Gaussian erasure channel. 
Specifically, it assumes $y=XX'x+z$,
(where $x$ and $y$ are the channel input and output)
followed by $n-k$ random erasures. 
The quantity $\alpha$ in \eqref{shannon_func:eq} is the 
signal-to-noise ratio $SNR=\alpha\geq 0$.
\\
~
\\
\noindent In this paper, we focus on {\em typical-case} performance criteria (those that
seek to optimize $\E_K \specstat(\lambda(\Gk))$ over random choice of $K$) rather than
{\em worse-case} performance criteria (those that seek to optimize 
$\max_{K\subset [n]} \specstat(\lambda(\Gk))$, such as RIP). For the remainder of this
paper, $K\subset[n]$ will denote a uniformly distributed random subset of size
$k$. Importantly, $k$ should be allowed to be large, even as large as $\m$. 

For a given $\specstat$, one would like to design frames that optimize $\E_K
\specstat(\lambda(\Gk))$.  This turns out to be a difficult task; in fact, it is
not even known how to calculate $\E_K \specstat(\lambda(\Gk))$ for a given frame
$X$.
Indeed, to calculate this quantity one effectively has to average $\specstat$
over  the spectrum $\lambda(\Gk)$ for all $\binom{n}{k}$ subsets $K\subset [n]$.
It is of little surprise to the information theorist that the first frame
designs, for which performance was formally bounded (and still not calculated
exactly), consisted of random vectors
\cite{CT06,Nelson,Rudelson,Oded,Tropp,Charaghchi}.


\section*{Random Frames} \label{sec:random_frames}

When the frame is random, namely when $X$ is drawn from some ensemble of random
matrices, the typical $k$-submatrix $\Xk$ is also a random matrix. Given a
specific $\specstat$, rather than seeking to bound $\E_K
\specstat(\lambda(\Gk))$ for specific $n$ and $\m$, it can be extremely
rewarding to study the limit of $\specstat(\lambda(\Gk))$ as the frame size $n$
and $\m$ grow. This is because tools from random matrix theory become available,
which allow exact asymptotic calculation of $\lambda(\Gk)$ and
$\specstat(\lambda(\Gk))$, and also because their limiting values are usually
very close to their corresponding values for finite $n$ and $\m$, even for low
values of $n$.

Let us consider then a sequence of dimensions $\m_n$ with $\m_n/n=\gamma_n\to \gamma$
and a sequence of 
random frame matrices $X^{(n)}\subset \R^{n\times\m_n}$ or $\C^{n\times\m_n}$.
To
characterize the collective behavior of $k$-submatrices we choose a sequence
$k_n$ with $k_n/\m_n=\beta_n\to \beta$ 
and look at the spectrum
$\lambda(\Gkn)$ 
of the random
matrix 
$\Xkn$ as $n\to\infty$, where $K_n\subset [n]$ is a randomly chosen 
subset with $|K_n|=k_n$. Here and below, to avoid cumbersome notation we omit the subscript $n$ and write $m$,$k$ and $K$ for $m_n$,$k_n$ and $K_n$.

A mainstay of random matrix theory is the celebrated convergence of the
empirical spectral distribution of random matrices, drawn from a certain
ensemble, to a limiting spectral distribution corresponding to that ensemble. 
This has indeed been established for three random frames:
\begin{enumerate} 

\item {\em Gaussian i.i.d frame:} Let $X_{normal}^{(n)}$ have i.i.d normal
entries with mean zero and variance $1/\m$.  The empirical
distribution of $\lambda(\Gk)$ famously converges, almost surely in
distribution, to the Mar\u cenko-Pastur density \cite{MP} with parameter
$\beta$:  
\begin{equation}
\label{MPdensity:eq}
f_{\beta}^{MP}(x)
 =\frac{\sqrt{(x-\lambda^{MP}_-)(\lambda^{MP}_+-x)}}{2\beta\pi x}\cdot
I_{(\lambda^{MP}_-,\lambda^{MP}_+)}(x),
\end{equation} 
supported on $[\lambda^{MP}_-,\lambda^{MP}_+]$ where
$\lambda^{MP}_\pm = (1\pm \sqrt{\beta})^2$.  Moreover,  almost surely
$\lambda_{max}(G_{normal}^{(n)}) \to \lambda_+$ and
$\lambda_{min}(G_{normal}^{(n)}) \to \lambda_-$; in other words, the maximal
and minimal empirical eigenvalues converge almost surely to the edges of the
support of the limiting spectral distribution \cite{silverstein_book}.

\item {\em Random Fourier frame:}
Consider the
random Fourier frame, in which the $\m_n$ columns of $X_{fourier}^{(n)}$ are drawn
uniformly
at random
from the columns of the $n$-by-$n$ discrete Fourier transform (DFT) matrix
(normalized s.t absolute value of matrix entries is $1/\sqrt{\m}$). 
Farrell \cite{Farrell} has proved that the 
empirical distribution of $\lambda(\Gk)$ converges, almost surely in
distribution,  as $n\to\infty$ and as $m$ and $k$ grow proportionally to $n$, 
to the so-called MANOVA
limiting distribution, which we now describe briefly.



%
The classical MANOVA$(n,\m,k,\Fc)$ ensemble\footnote{Also known as the
beta-Jacobi ensemble with beta=$1$ (orthogonal) 
for $\Fc=\R$, and beta=$2$ (unitary) for $\Fc=\C$.}, with $\Fc\in\left\{ \R,\C
\right\}$ is the distribution of the random 
matrix 
\begin{equation}
\label{ManovaRandomMatrix}
\frac{n}{\m}(AA'+BB')^{-\frac{1}{2}}BB'(AA'+BB')^{-\frac{1}{2}}\,,
\end{equation} 
where $A_{k\x (n-\m)}, B_{k\x \m}$ are random standard Gaussian
i.i.d matrices with entries in $\Fc$.  
Wachter \cite{wachter} discovered that, as $k/\m\to \beta \le 1$ and $\m/n\to\gamma$, 
the empirical spectral
distribution of the MANOVA$(n,\m,k,\R)$ ensemble converges, almost surely in distribution,
to the so-called
MANOVA$(\beta,\gamma)$ limiting
spectral distribution\footnote{The literature uses the term MANOVA to refer both
	to the random matrix ensemble, which we denote here by MANOVA$(n,\m,k,\Fc)$,
	and to the limiting spectral distribution, which we denote here by
MANOVA$(\beta,\gamma)$.}, whose density is given by 
%
\begin{eqnarray}
\label{ManovaDensity}
f_{\beta,\gamma}^{MANOVA}(x)
 =\frac{\sqrt{(x-r_-)(r_+-x)}}{2\beta\pi x(1-\gamma x)}\cdot
I_{(r_-,r_+)}(x) & & \\
   + \left(1+\frac{1}{\beta}-\frac{1}{\beta\gamma}\right)^+ \cdot \delta\left(x-\frac{1}{\gamma}\right) \, &&
   \nonumber
\end{eqnarray}
where $(x)^+ = \max(0,x)$.
The limiting MANOVA distribution is compactly supported on $[r_-,r_+]$  with
\begin{equation}
\label{ManovaDensityExtrimalValues}
r_\pm=\bigg(\sqrt{\beta(1-\gamma)}\pm\sqrt{1-\beta\gamma}\bigg)^2\,.
\end{equation} 
\noindent The same holds for the MANOVA$(n,\m,k,\C)$ ensemble. 

Note that the support of the MANOVA$(\beta,\gamma)$ distribution is smaller than
that of the corresponding Mar\u cenko-Pastur law for the same aspect ratios. Figure
\ref{fig:MANOVA_MP} shows these two densities for $\beta=0.8$ 
and $\gamma=0.5$.
Nevertheless, as the MANOVA dimension ratio becomes small,
its distribution tends to the Mar\u cenko-Pastur distribution \eqref{MPdensity:eq},
i.e., $f^{MANOVA}_{\beta,\gamma}(x) \rightarrow f^{MP}_\beta(x)$ as $\gamma \rightarrow 0$.
Thus, a highly redundant random Fourier frame behaves like
a Gaussian i.i.d. frame.
\begin{figure}[h]
\centering
\includegraphics[width=5in]{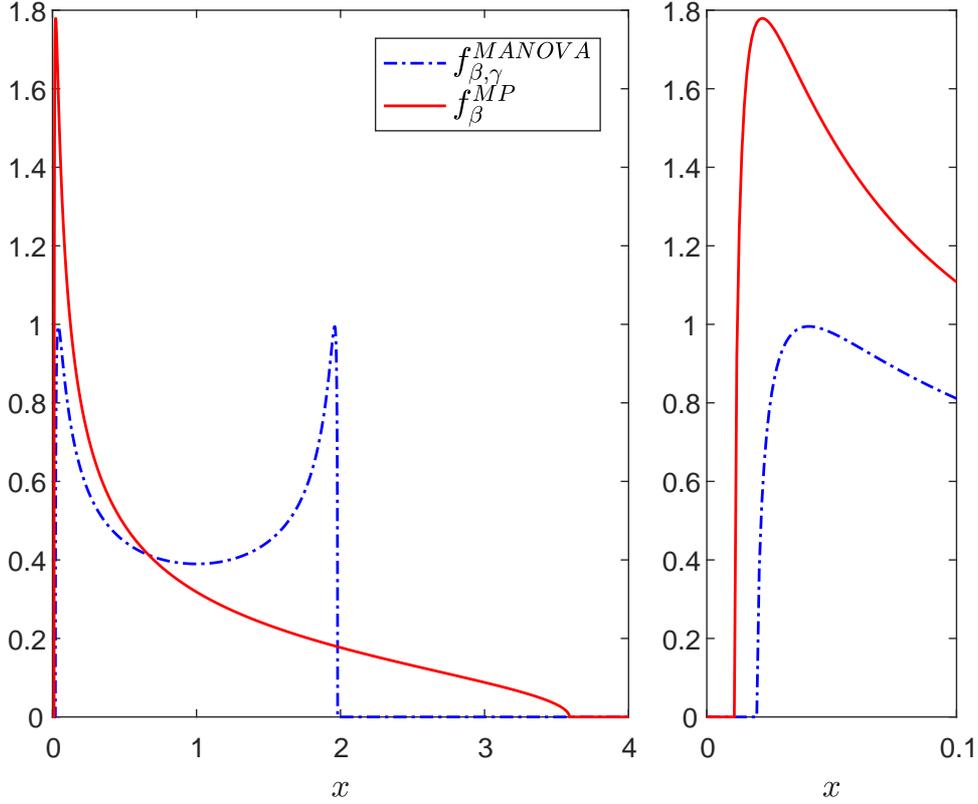}
\caption{Limiting 
	MANOVA ($\beta=0.8,\gamma=0.5$) 
and Mar\u cenko-Pastur ($\beta=0.8$) density
functions. Left: density on the interval $x\in [0,4]$. Right: Zoom in on the
interval $x\in[0,0.1]$. 
}
\label{fig:MANOVA_MP}
\end{figure}

\item {\em Unitary Haar frame:}
Let $X_{haar}^{(n)}$ consist of
the first $\m$ columns of a
Haar-distributed $n$-by-$n$ unitary matrix normalized by $\sqrt{n/\m}$ (the Haar distribution being the
uniform distribution over the group of $n$-by-$n$ unitary matrices).
Edelman and Sutton \cite{Edelman} proved that 
the empirical spectral distribution of $\lambda(\Gk$) also converges, almost
surely in distribution, to the
MANOVA limiting spectral distribution 
(See also \cite{wachter} and the closing remarks of \cite{Farrell}.)
\end{enumerate}
The maximal and minimal eigenvalues of a matrix
from the MANOVA$(n,\m,k,\Fc)$ ensemble ($\Fc\in\left\{ \R,\C \right\}$) 
are known to converge almost surely 
to $r_+$ and $r_-$, respectively \cite{Johnstone2008}.
While we are not aware of any parallel results for the random Fourier and Haar
frames,  
the empirical evidence in this paper show that it must be the case.

These random matrix phenomena have practical significance for evaluations of
functions of the form $\specstat(\lambda(\Gk))$ such as those mentioned above.
The functions $\specstat_{AC}$ and $\specstat_{Shannon}$, for example, are what \cite{BaiBook}
call {\em linear spectral statistics}, namely functions of $\lambda(\Gk)$ that
may be written as an integral of a scalar function against the empirical measure of $\lambda(\Gk)$.
Convergence of the empirical
distribution of $\lambda(\Gk^{(n)})$ to the limiting 
MANOVA distribution with density $f^{MANOVA}_{\beta,\gamma}$ 
implies 
\begin{eqnarray} \label{ac_limit:eq}
\lim_{n\to\infty} \specstat_{AC}(\lambda(\Gkn^{(n)})) &=& \int \frac{1}{x}
\,f_{\beta,\gamma}^{MANOVA}(x)dx \\
\lim_{n\to\infty} \specstat_{Shannon}(\lambda(\Gkn^{(n)})) &=&  \int \log(1+\alpha x)
\,f_{\beta,\gamma}^{MANOVA}(x)dx \nonumber 
\end{eqnarray}
for both the random Fourier and Haar frames; 
the integrals on the right hand side may be evaluated explicitly.
Similarly, convergence of 
$\lambda_{max}(\Gk)$ and $\lambda_{min}(\Gk)$ to $r_+$ and $r_-$ implies, for
example, that 
\begin{eqnarray} \label{rip_limit:eq}
\lim_{n\to\infty} \specstat_{RIP}(\lambda(\Gk^{(n)})) = \max(r_+-1,1-r_-)\,.
\end{eqnarray}

\begin{figure*}[h!]
\centering
\includegraphics[width=3in]{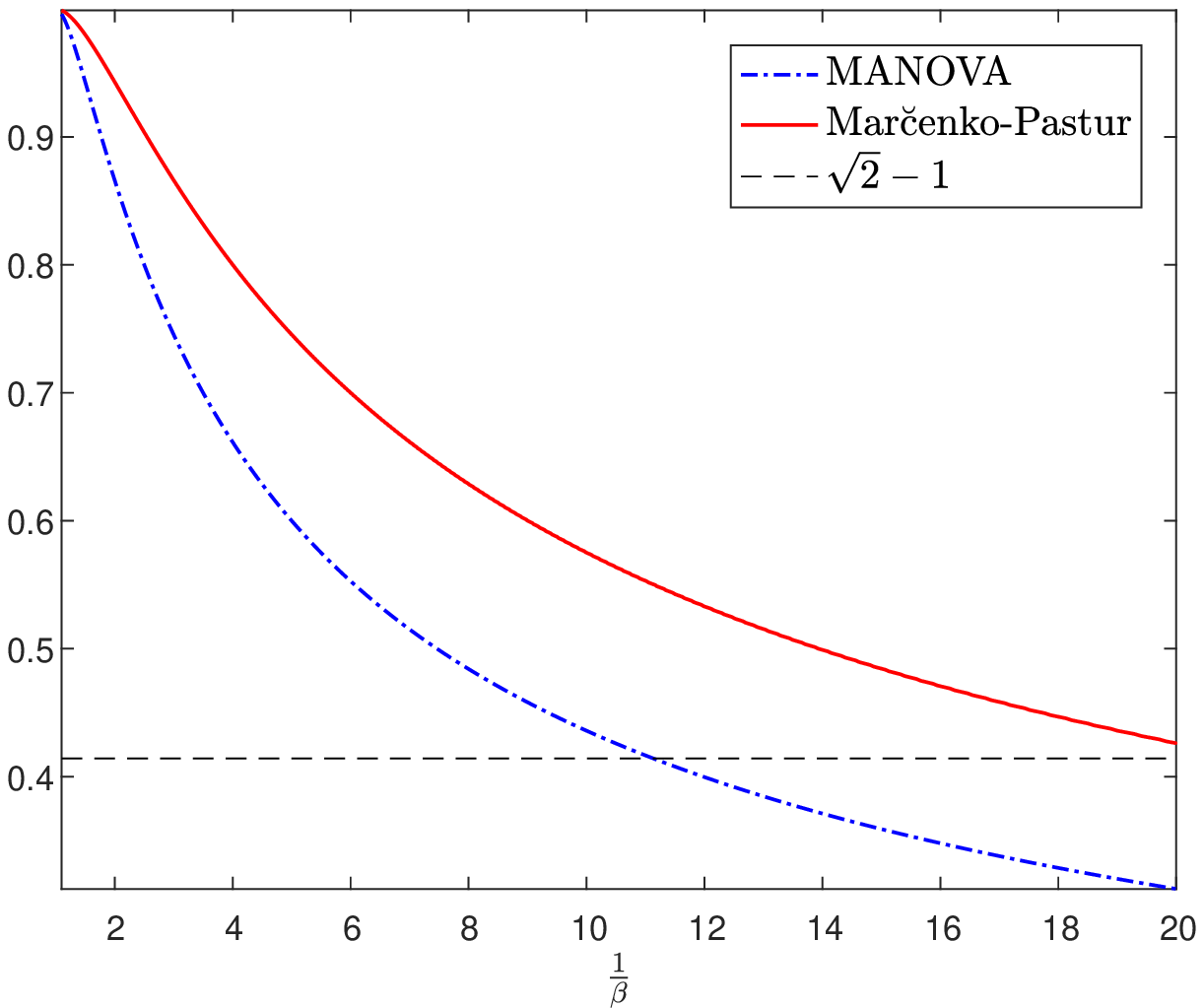}\\
\includegraphics[width=3in]{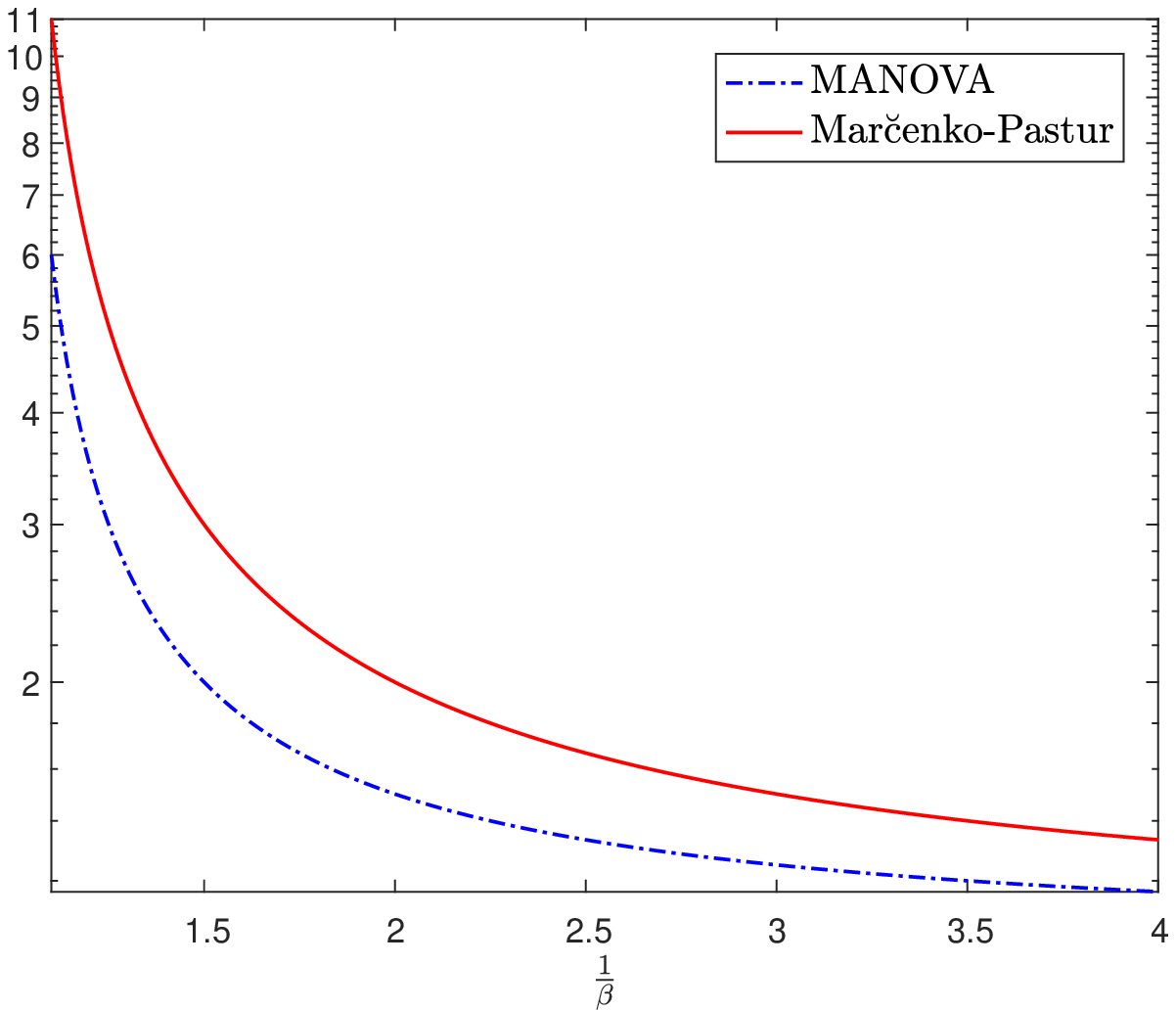}\\
\includegraphics[width=3in]{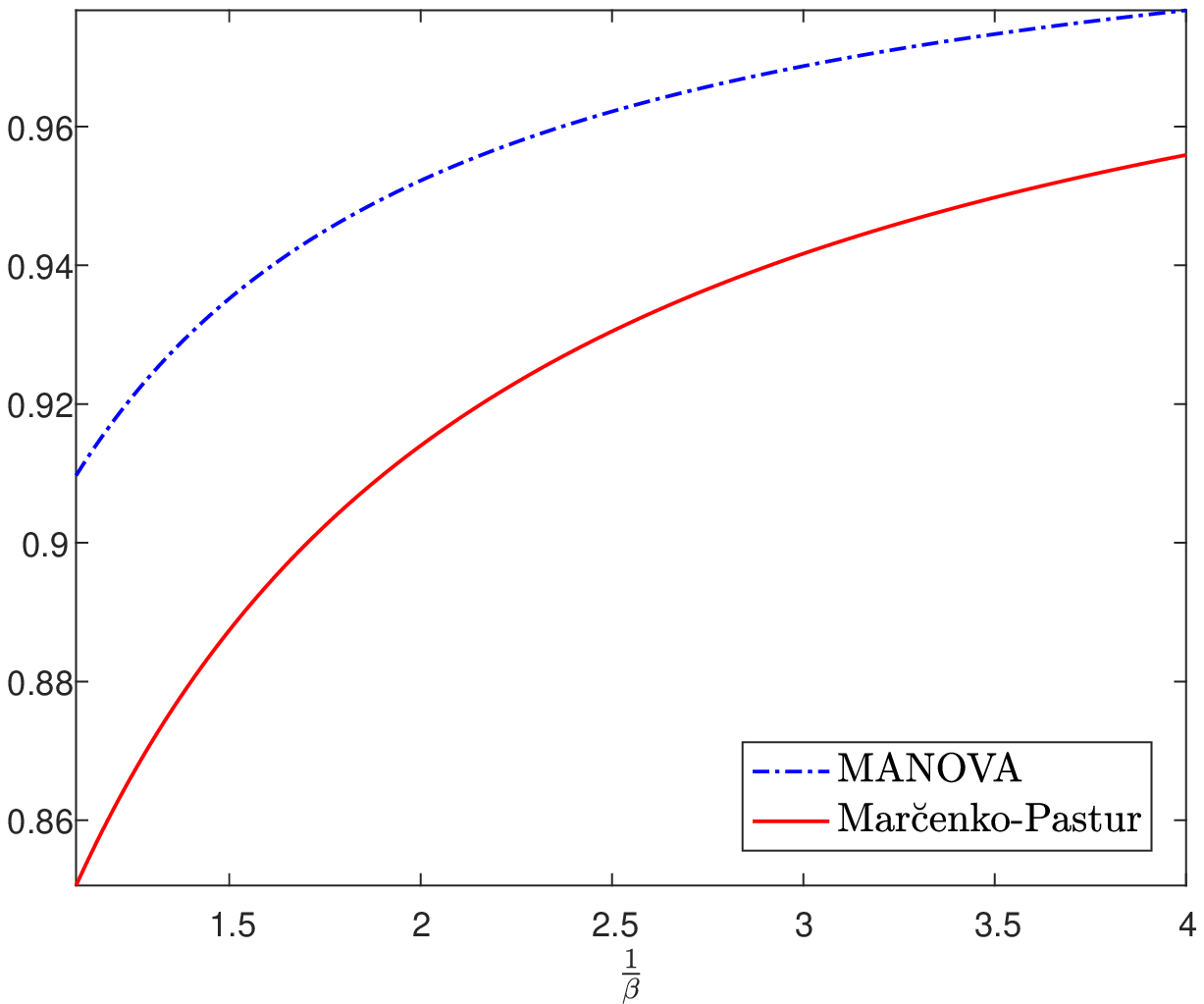}
\caption{Comparison of limiting values of $\E_K \specstat(\lambda(G_K))$ for the three
functions $\specstat$ discussed in Motivation Section between the Mar\u cenko-Pastur limiting distribution and the MANOVA
distribution. Left: $\specstat_{RIP}$ (lower is better). 
Middle: $\specstat_{AC}$ (lower is
better). Right: $\specstat_{Shannon}$ (higher is better).
}
\label{fig:limiting_F}
\end{figure*}

To demonstrate why such calculations are significant,
we note that Equations \eqref{ac_limit:eq} and \eqref{rip_limit:eq} 
immediately allow us to compare the Gaussian i.i.d frame
with the random Fourier and Haar frames, 
in terms
of their limiting value of functions of interest.
Figure \ref{fig:limiting_F} compares the limiting value of $\specstat_{RIP}$,
$\specstat_{AC}$ and $\specstat_{Shannon}$ over varying values of $\beta=\lim_{n\to \infty}
k/\m$. The plots 
clearly demonstrate that frames whose typical
$k$-submatrix exhibits a MANOVA spectrum, are superior to frames whose typical
$k$-submatrix exhibits a Mar\u cenko-Pastur spectrum, 
across the performance measures.

%
%
%
%
%


\section*{Deterministic Frames: Universality Hypothesis}
\label{sec:deterministic_frames}

Deterministic frames, namely
frames whose design involves no randomness, have so far eluded this kind of
asymptotically exact analysis. 
While
there are results regarding RIP \cite{Bandiera,Fickus} and statistical RIP
\cite{Caulderbank-STRIP,Gurevich2009,Mazumdar}, for example, of deterministic
frame designs, they are mostly focused on highly redundant frames ($\gamma \rightarrow 0$)
and the wide submatrix ($\beta \rightarrow 0$) case, where the spectrum tends
to the Mar\u cenko-Pastur distribution. Furthermore, nothing analogous, say, to the precise comparisons of Figure
\ref{fig:limiting_F} exists in the literature to the best of our knowledge. 
Specifically, no results analogous to \eqref{ac_limit:eq} and \eqref{rip_limit:eq}
are known for deterministic frames, 
let alone the associated convergence rates, if any.

In order to subject deterministic frames to an asymptotic analysis, we shift our
focus from a single frame $X$ to a family of deterministic frames $\{X^{(n)}\}$
created by a common construction. The frame matrix $X^{(n)}$ is $n$-by-$m$.
Each frame family determines allowable sub-sequences $(n,m)$; to
simplify notation, we leave the subsequence implicit and index the frame
sequence simply by $n$. The frame family
also determines the aspect ratio limit $\gamma=\lim_{n\to\infty} \m/n$.
In what follows we also fix a sequence $k$ with $\beta=\lim_{n\to\infty}
k/\m $,
and let $K\subset [n]$ denote a uniformly distributed random subset.

\paragraph{Frames under study.} 

The different frames that we studied are listed in Table \ref{FramesTable},
in a manner inspired by \cite{Monajemi}.
In addition to our deterministic frames of interest 
(the set ${\cal X}$),
the table contains also two examples of random frames
(real and complex variant for each), 
for validation and convergence analysis purposes.
%
\begin{table*}[t]
\centering
\caption{Frames under study}
\label{FramesTable}
{\scriptsize
\begin{tabular}{lllllll}
\toprule
Label & Name & $\R$ or $\C$ & Natural $\gamma$ & Tight frame &
Equiangular & References \\
\midrule
& & & & &  & \\
{\bf Deterministic frames} & & & & &  & \\
DSS & Difference-set spectrum & $\C$  & & Yes & Yes & \cite{WBdss}\\
GF & Grassmannian frame & $\C$ & $1/2$& Yes & Yes &  \cite[Cor. 2.6b]{Grassmannian}
\\
RealPF & Real Paley's construction & $\R$ & $1/2$ & Yes & Yes & \cite[Cor. 2.6a]{Grassmannian}
\\
ComplexPF & Complex Paley's construction & $\C$ & $1/2$ & Yes & Yes & 
\cite{PaleyConstruction}
\\
Alltop & Quadratic Phase Chirp & $\C$ & $1/L$ & Yes & No &  
\cite[eq. S4]{Monajemi} with $L=2$
\\
SS & Spikes and Sines & $\C$ & $1/2$ & Yes & No & \cite{Elad2010} \\
SH & Spikes and Hadamard & $\R$  & $1/2$ & Yes & No & \cite{Elad2010}
~ \\\hline  ~\\
{\bf Random frames} & & & & &  & \\
HAAR 
& Unitary Haar frame & $\C$ & & Yes & No & \cite{Farrell,Edelman} \\
RealHAAR 
& Orthogonal Haar frame & $\R$ & & Yes & No & \cite{Edelman} \\
RandDFT  
& Random Fourier transform & $\C$ & & Yes & No & \cite{Farrell} \\
RandDCT  
& Random Cosine transform & $\R$ & & Yes & No &  \\

\bottomrule
\end{tabular}
}
\label{frames:tab}
\end{table*}


\paragraph{Functionals under study.}

We studied the functionals $\specstat_{StRIP}$ from \eqref{strip_func:eq}, 
$\specstat_{AC}$ from \eqref{ac_func:eq}, $\specstat_{Shannon}$ from
\eqref{shannon_func:eq}. In addition, we studied the maximal and minimal
eigenvalues of $\Gk$, and its condition number:
\begin{eqnarray*}
\specstat_{max}(\lambda(\Gk)) &=&  \lambda_{max}(\Gk) \\
\specstat_{min}(\lambda(\Gk)) &=&  \lambda_{min}(\Gk) \\
\specstat_{cond}(\lambda(\Gk)) &=&  \lambda_{max}(\Gk) / \lambda_{min}(\Gk)\,. 
\end{eqnarray*}

\begin{figure}[h]
\centering
\includegraphics[width=5in]
{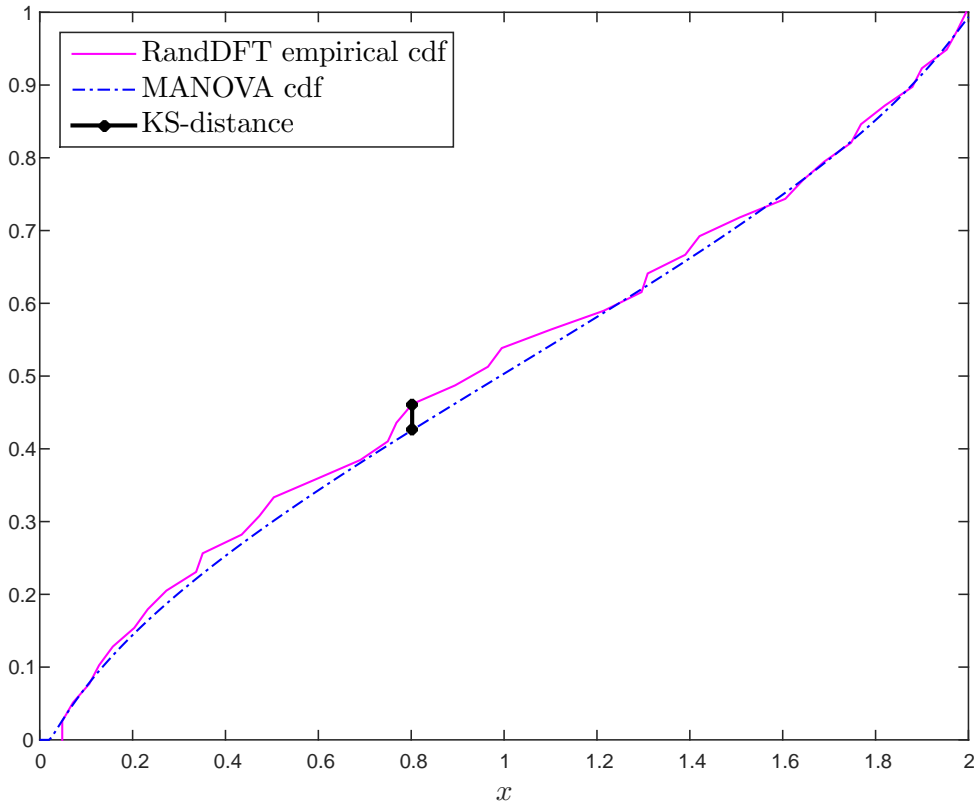}
\caption{KS-distance of random DFT subframe, $\beta = 0.8$, $\gamma = 0.5$, $n=100$.}
\label{fig:CDFs}
\end{figure}

\paragraph{Measuring the rate of convergence.}
In order to quantify the rate of convergence of the entire spectrum 
of the $k$-by-$\m$ matrix 
$\Xk$, which is a $k$-submatrix of an $n$-by-$\m$ frame matrix $X$, to a limiting
distribution, we let $F[\Xk]$ denote the
empirical cumulative distribution function (CDF) of $\lambda(\Gk)$, and 
let $F^{MANOVA}_{\beta,\gamma}(x) = \intop_{r_-}^x
f^{MANOVA}_{\beta,\gamma}(x)dx$ denote the CDF of the MANOVA$(\beta,\gamma)$
limiting distribution. 
The quantity
\[
\Delta_{KS}(\Xk) = \norm{F[\Xk] - F^{MANOVA}_{\beta_n,\gamma_n}}_{KS}\,,
\]
where $\norm{\cdot}_{KS}$ is the Kolmogorov-Smirnov (KS) distance between CDFs, 
measures the distance to the hypothesised limit. Here, 
$\beta_n=k/m$ and $\gamma_n=m/n$ are the actual aspect ratios for the matrix $\Xk$ at hand.
As a baseline we use $\Delta_{KS}( Y_{n,\m,k,\Fc})$, where 
$Y_{n,\m,k,\Fc}$ is a matrix from the MANOVA$(n,\m,k,\Fc)$ ensemble,
with $\Fc=\R$ if $X_K$ is real and $\Fc=\C$ if complex.
Figure \ref{fig:CDFs} illustrates the KS-distance 
between an empirical CDF and the limiting MANOVA CDF.

Similarly, in order to quantify the rate of convergence of a functional
$\specstat$, the quantity
\[
\Delta_\specstat(\Xk;n,\m,k) = \big| \Psi(\lambda(\Gk)) -
\Psi(f^{MANOVA}_{\beta_n,\gamma_n}) \big|
\]
is the distance between the measured value of $\Psi$ on a given $k$-submatrix
$\Xk$ and its hypothesised limiting value. 
For a baseline we can use $\Delta_\specstat(Y_{n,\m,k,\Fc})$,
with $\Fc=\R$ if $X_K$ is real and $\Fc=\C$ if complex.
For linear spectral functionals 
like $\Psi_{AC}$ and $\Psi_{Shannon}$, which may be written as
$\Psi(\lambda(\Gk))=\int \psi dF[X_K]$ for some kernel $\psi$, we have
$  \Psi(f^{MANOVA}_{\beta,\gamma}) = \int \psi dF^{MANOVA}_{\beta,\gamma}$. 
For $\Psi_{RIP}$ that depends on $\lambda_{max}(\Gk)$ 
and $\lambda_{min}(\Gk)$ we have $\Psi_{RIP}(f^{MANOVA}_{\beta,\gamma}) =
\max\left\{ r_+-1,1-r_- \right\}$.

\paragraph{Universality Hypothesis.} 

The contributions of this paper are based on the following assertions on the typical $k$-submatrix ensemble $\Xk$ corresponding to a frame
family $X^{(n)}$. This family may be random or deterministic, real or complex. 

\begin{enumerate}
\item[{\bf H1}] {\em Existence of a limiting spectral distribution.} 
The empirical spectral distribution of $\Xk^{(n)}$, namely the 
distribution of $\lambda(\Gk^{(n)})$, converges, as $n\to \infty$, 
to a compactly-supported limiting distribution; furthermore, 
$\lambda_{max}(\Gk^{(n)})$ and $\lambda_{min}(\Gk^{(n)})$ converge to the
edges of that compact support.

\item[{\bf H2}] {\em Universality of the limiting spectral distribution.} 
The limiting 
spectral distribution of $\Xk^{(n)}$ is the 
MANOVA$(\beta,\gamma)$ distribution  \cite{wachter} whose density is 
\eqref{ManovaDensity}. Also 
$\lambda_{max}(\Gk^{(n)})\to r_+$ and $\lambda_{min}(\Gk^{(n)})\to r_-$
where $r_\pm$ is given by \eqref{ManovaDensityExtrimalValues}.


\item[{\bf H3}] {\em Exact power-law rate of convergence for the entire
spectrum.}
The spectrum of $\Xk^{(n)}$ converges to the limiting
MANOVA$(\beta,\gamma)$ distribution
\begin{eqnarray*} 
   \left(\E_{K_n}\left(\Delta_{KS}(\Xk^{(n)})\right)\right)^2 \searrow 0
\end{eqnarray*}
and in fact its fluctuations are given by the law
\begin{eqnarray} \label{KS_conv:eq}
Var_{K}(\Delta_{KS}(\Xk^{(n)}))=Cn^{-2b} 
\end{eqnarray}
for some constants $C,b$, which may depend on the frame family.

\item[{\bf H4}] {\em Universality of the rate of convergence for the
entire spectrum of ETFs.}
For an equiangular tight frame (ETF) family,  
the exponent
$b$ in \eqref{KS_conv:eq} is universal and does not depend on the frame. 
Furthermore, 
\eqref{KS_conv:eq} also holds, with the same universal exponent,
replacing $\Gk^{(n)}$ with a same-sized matrix
from the MANOVA$(n,\m,k,\Fc)$ distribution 
defined in \eqref{ManovaRandomMatrix}, with $\Fc=\R$ if $X^{(n)}$ is a real
frame family, and $\Fc=\C$ if complex. 
In other words, the universal exponent $b$ for ETFs 
is a property of the MANOVA
(Jacobi) random matrix ensemble.


\item[{\bf H5}] {\em Exact power-law rate of convergence for functionals.}
For a ``nice'' functional $\specstat$, the value of
$\specstat(\lambda(\Gk^{(n)}))$  converges
to $\specstat(f^{MANOVA}_{\beta,\gamma})$ according to the law
\begin{eqnarray} \label{func_conv:eq}
\E_{K}(\Delta_\specstat(\Xk^{(n)})^2)=Cn^{-b}\log^{-a}(n)
\end{eqnarray}
for some constants $C,b,a$.

\item[{\bf H6}]  {\em Universality of the rate of convergence for functionals.}
While the constant $C$ in \eqref{func_conv:eq} may depend on the frame,
the exponents 
$a,b$ are universal. \eqref{func_conv:eq} 
also holds, with the same universal exponents, 
replacing $\Gk^{(n)}$ with a same-sized matrix 
from the MANOVA$(n,\m,k,\Fc)$ ensemble 
defined in \eqref{ManovaRandomMatrix}, with  
$\Fc=\R$ if $X^{(n)}$ is a real
frame family, and $\Fc=\C$ if complex. 
In other words, the universal exponents $a,b$ 
are a property of the MANOVA
(Jacobi) random matrix ensemble.


\end{enumerate}

\paragraph{Nonstandard aspect ratio $\beta>1$.}
While the classical MANOVA ensemble and limiting density are not defined for
$\beta>1$, in our case it is certainly possible to sample $k>m$ vectors from the
$n$ possible frame vectors, resulting in a situation with $\beta>1$.
In this situation, the hypotheses above require slight modifications.
Specifically, the limiting 
spectral distribution of $\Xk^{(n)}$, for $\beta>1$, is
\begin{equation}
\label{ManovaDensityBeta}
	\left(1-\frac{1}{\beta}\right) \delta(x)+f_{\beta,\gamma}^{MANOVA}(x)\,,
\end{equation}
where $f_{\beta,\gamma}^{MANOVA}(x)$ is the function (no longer a density) 
defined in \eqref{ManovaDensity}.
The rate of convergence of the distribution of nonzero eigenvalues to the
limiting density
$\frac{1}{\beta}f_{\frac{1}{\beta},\beta\gamma}^{MANOVA}(\frac{1}{\beta}x)=\beta
f_{\beta,\gamma}^{MANOVA}(x)$ is compared with 
the baseline $\beta\cdot Y_{n,k,\m,\Fc}$, where $Y_{n,k,\m,\Fc}$ is a matrix from the MANOVA$(n,k,\m,\Fc)$ ensemble
(i.e., with reversed order of $k$ and $m$).



\section*{Methods} \label{sec:methods}

The software we developed has been permanently deposited 
in the Data and Code Supplement \cite{SDR}.
As many of the deterministic frames under study are only defined for $\gamma=0.5$, 
we primarily studied the aspect ratios $(\gamma=0.5,\beta)$ with 
$\beta\in\{0.3,0.5,0.6,0.7,0.8,0.9\}$.
In addition, we inspected all frames under study that are defined for the aspect ratios $(\gamma=0.25,\beta=0.6)$ and $(\gamma=0.25,\beta=0.8)$ (all random frames, as well as DSS and Alltop).
We also studied nonstandard aspect ratios $\beta>1$ as described in the
Supporting
Information \cite{SI}. 
For deterministic frames, $n$ took allowed values in the range 
$(240,2000)$, $(2^5,2^{12})$ for Grassmannian and Spikes and Hadamard frames and $(600,4000)$ for DSS frame with $\gamma=0.25$.
For random frames and MANOVA ensemble we used
dense grid of values in the range $(240,2000)$. 
Hypothesis testing as discussed below, 
was based on a subset of these values where $n\ge 1000$.
For each of the frame families under study, and for each value of $\beta$ and
$\gamma$ under study, we selected a sequence $(n,\m,k)$.
The values $n$ and $m$ were selected so that $m/n$ will be as close as
possible to $\gamma$, however due to different aspect ratio constrains by the
different frames occasionally we had $m/n$ close but not equal to $\gamma$.
We then determined $k$ such that $k/\m$ will be as close as
possible to $\beta$.
For each $n$, we generated a single 
$n$-by-$\m$ frame matrix
$X^{(n)}$.
We then produced $T$ independent samples from the uniform
distribution on $k_n$-subsets,
$K[1],\ldots,K[T]\subset[n]$, and generated 
their corresponding $k$-submatrices 
$X_{K[i]}^{(n)}$ ($1\leq i\leq T$). Importantly, all these are submatrices of
the same frame matrix $X^{(n)}$.
We calculated $\overline{\Delta}^{Var}_{KS}(\Xk^{(n)})=\overline{\Delta^2}_{KS}(\Xk^{(n)})-\overline{\Delta}^2_{KS}(\Xk^{(n)})$, the empirical variance of 
$ \Delta_{KS}(X_{K[i]}^{(n)})$, and $\overline{\Delta^2}_{KS}(\Xk^{(n)})$, the average value of $ \Delta^2_{KS}(X_{K[i]}^{(n)})$ on $1\leq i\leq T$, as a 
monte-carlo approximation  to the left-hand side of \eqref{KS_conv:eq}, variance and MSE respectively. For each
of the functionals under study, we also calculated 
$ \overline{\Delta^2}_\specstat(X_{K[i]}^{(n)})$,
the average value of 
$ \Delta^2_\specstat(X_{K[i]}^{(n)})$ on 
$1\leq i\leq T$, as a monte-carlo approximation to the left-hand size of 
\eqref{func_conv:eq}.

Separately, 
for each triplet ($n$,$m$,$k$) and $\Fc\in\left\{ \R,\C \right\}$
we have performed $T$ independent draws from
the MANOVA$(n,\m,k,\Fc)$ ensembles 
(\ref{ManovaRandomMatrix}) and calculated 
analogous quantities 
$\overline{\Delta}^{Var}_{KS}(Y_{n,\m,k,\Fc})$,
$\overline{\Delta^2}_{KS}(Y_{n,\m,k,\Fc})$ 
and
$\overline{\Delta^2}_{\specstat}(Y_{n,\m,k,\Fc})$.

\paragraph{Test 1: Testing H1--H4.}
For each of the frames under study and each value of $(\beta,\gamma)$, we computed the KS-distance for $T=10^4$
submatrices and performed simple linear regression
of 
$-\frac{1}{2}\log\left( \overline{\Delta}^{Var}_{KS}(\Xk^{(n)})\right)$ 
on $\log(n)$ with an intercept. We obtained 
the estimated linear coefficient $\hat{b}$ as an estimate
for the exponent $b$, and its standard error $\sigma(\hat{b})$.
Similarly we regressed 
$-\frac{1}{2}\log\left( 
\overline{\Delta}^{Var}_{KS}(Y_{n,\m,k,\Fc})\right)$
on $\log(n)$ to obtain $\hat{b}_{MANOVA}$ and $\sigma(\hat{b}_{MANOVA})$.  
We performed Student's
t-test to test the null hypotheses $b=b_{MANOVA}$ using
the test
statistic 
\[
t = \frac{\hat{b}-b_{MANOVA}}
{\sqrt{\sigma(\hat{b})^2+\sigma(\hat{b}_{MANOVA})^2 }}\,.
\]
Under the null hypothesis, 
the test statistic is distributed $t_{(N+N_{MANOVA}-4)}$, 
where $N$, $N_{MANOVA}$ are the numbers of different values of $n$ for which we have collected
the data for a frame and the MANOVA ensemble respectively.
We report the $R^2$ of
the linear fit; the slope coefficient $\hat{b}$ and its standard error; and the
p-value of the above t-test.
We next regressed $-\log \left(\overline{\Delta^2}_{KS}\right)$ on $\log(n)$. 
Since $\overline{\Delta^2}_{KS} = \left( \overline{\Delta}_{KS}
\right)^2+\overline{\Delta}_{KS}^{Var}$, a linear fit verifies that 
$ \left( \overline{\Delta}_{KS}
\right)^2 \searrow 0$.

\paragraph{Test 2: Testing H5--H6.}

For each of the frames under study, each of the functionals $\specstat$ 
under study, and 
each value of $(\beta,\gamma)$, we computed the empirical value of the
functionals on $T=10^3$ submatrices.
We first performed linear regression
%
%
of 
$-\log\left( \overline{\Delta^2}_{\specstat}(Y_{n,\m,k,\Fc})\right)$
on $\log(n)$ and $\log(\log(n))$ with an intercept, for $\Fc\in\left\{ \R,\C
\right\}$. 
Let $a_0$ denote the fitted coefficient for $\log(n)$ and let $b_0$ denote the
fitted coefficient for $\log(\log(n))$. 
This step was based on triplets $(n,m,k)$ yielding accurate aspect ratios in
the range $240\le n\le 2000$.
We then performed simple linear regression
of 
$-\log\left( \overline{\Delta^2}_{\specstat}(\Xk^{(n)};n,\m,k)\right)$
on 
$\log(n) + (a_0/b_0)\cdot \log(\log(n))$.
The estimated linear regression coefficient $\hat{b}$
is the estimate 
for the exponent $b$ in \eqref{func_conv:eq}, 
and $\sigma(\hat{b})$
is its standard error. We used 
$\hat{b}\cdot(a_0/b_0)$ as an
estimate for the exponent $a$ in
\eqref{func_conv:eq}.
We proceeded as above to test the null hypothesis $b=b_0$.
We report the $R^2$ of
the linear fit; the slope coefficient $\hat{b}$ and its standard error; and the
p-value of the test above.

\paragraph{Computing.}

To allow the number of monte-carlo samples to be as large as $T=10^4$ 
and $n$ to be as large as $2000$, we used a large Matlab cluster 
running on Amazon Web Services.
We used 32-logical core machines, with 240GB RAM each, 
which were running several hundred hours in total. 
The code we executed has been
deposited \cite{SDR}; it
may easily be executed for smaller values of $T$ and $n$ on smaller machines.



\section*{Results} \label{sec:results}

The raw results obtained in our experiments, as well as the analysis results 
of each experiment, have been deposited with their generating code \cite{SDR}.

\begin{figure}[h]
\centering
\includegraphics[width=5in]
{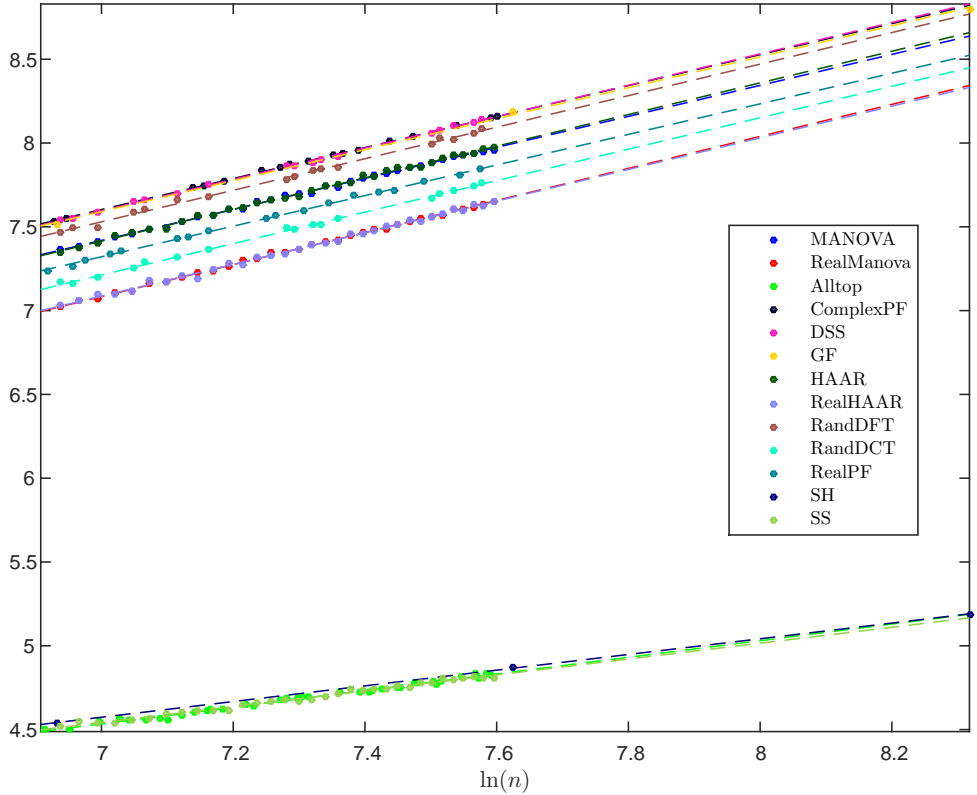}
\caption{Test 1 for $\gamma=0.5$ and $\beta=0.8$. Plot shows $-\frac{1}{2}\ln Var_{K}(\Delta_{KS}(\Xk^{(n)}))$ over $\ln(n)$. }
\label{fig:KS1}
\end{figure}

\begin{figure}[h]
\centering
\includegraphics[width=5in]{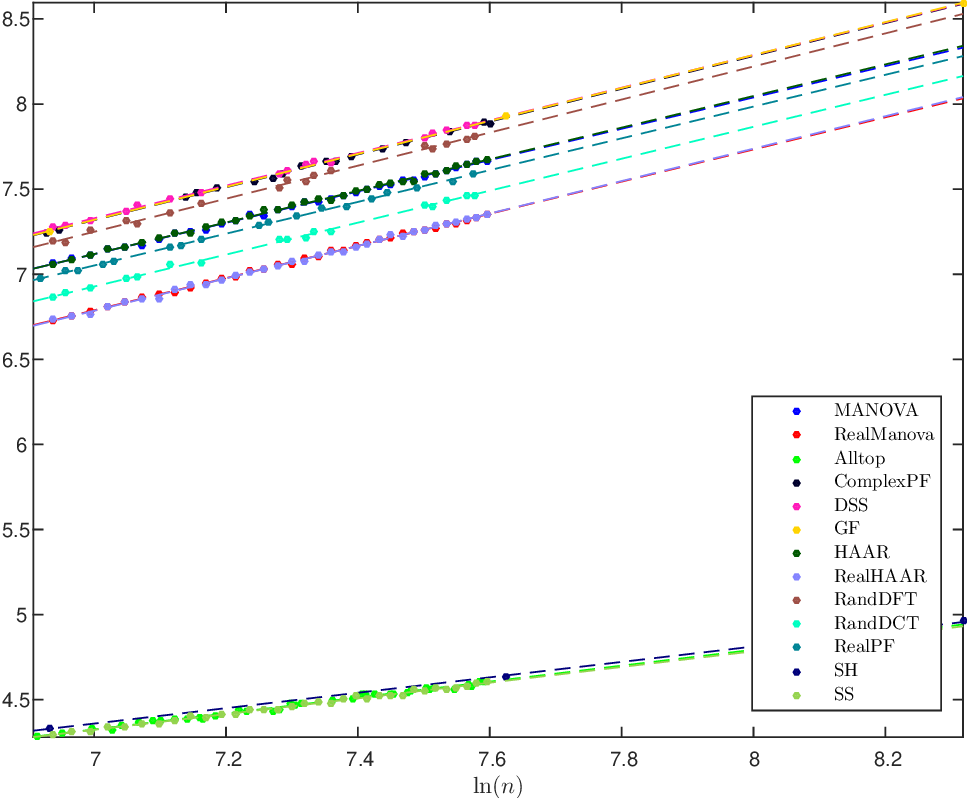}
\caption{Test 1 for $\gamma=0.5$ and $\beta=0.6$. Plot shows $-\frac{1}{2}\ln Var_{K}(\Delta_{KS}(\Xk^{(n)}))$ over $\ln(n)$. }
\label{fig:KS2}
\end{figure}

\begin{figure}[h]
\centering
\includegraphics[width=5in]{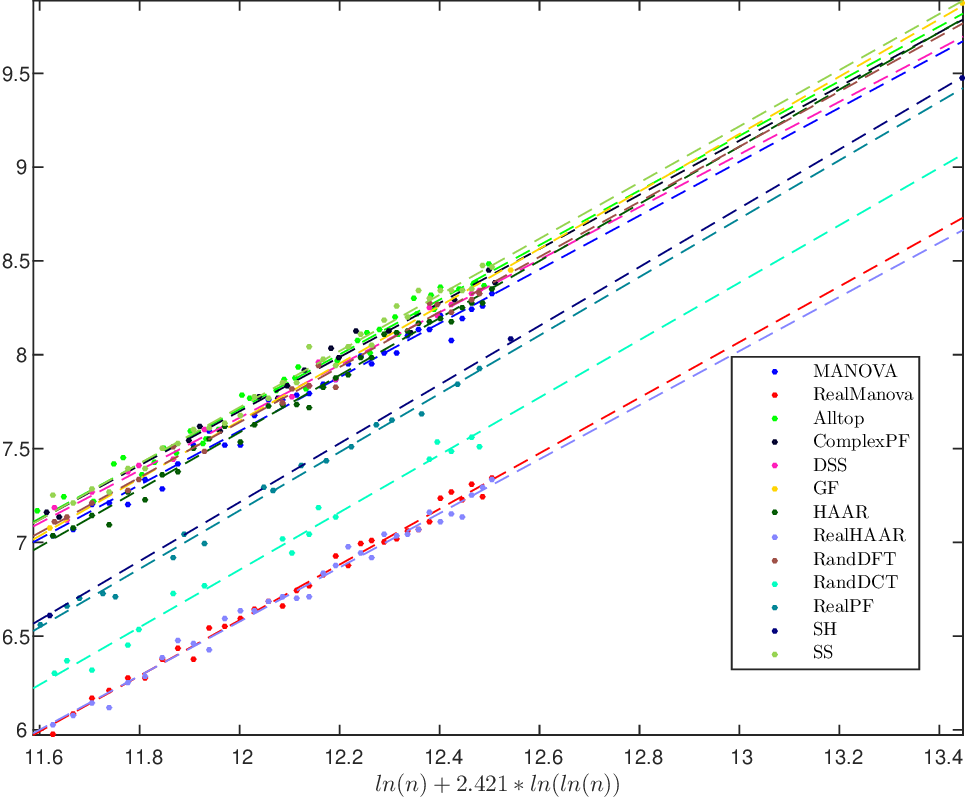}
\caption{Test 2 for $\specstat_{AC}$, $\gamma=0.5$ and $\beta=0.8$. Plot shows $-\ln\E_{K}(\Delta_\specstat(\Xk^{(n)})^2)$
}
\label{fig:f_AC}
\end{figure}

\begin{figure}[h]
\centering
\includegraphics[width=5in]{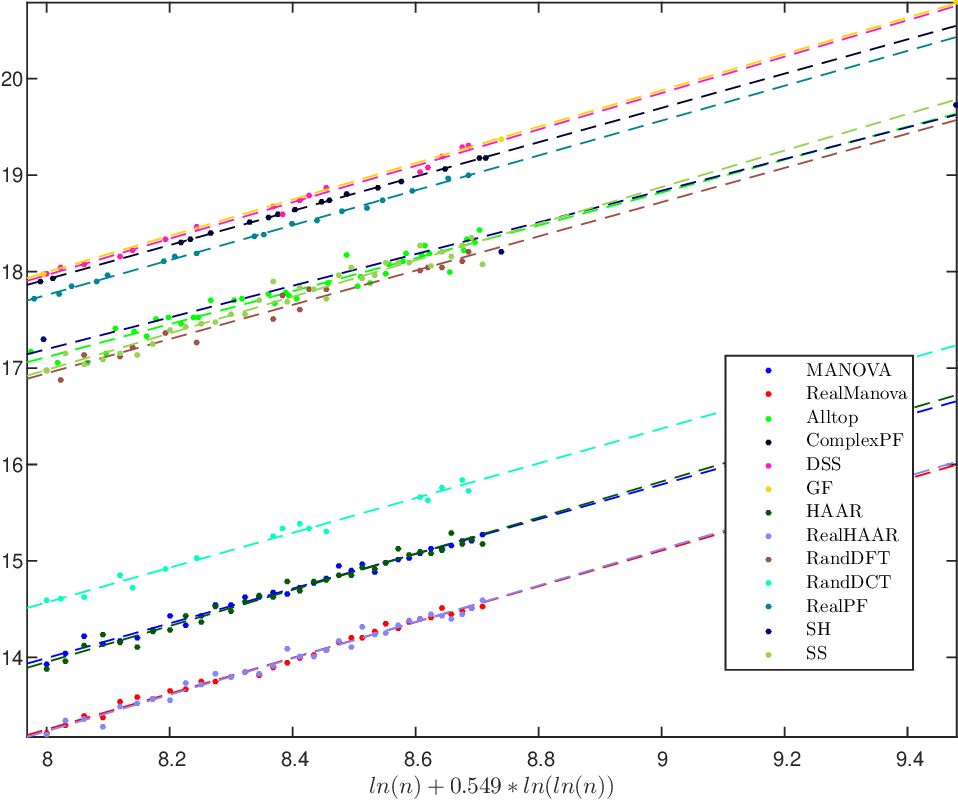}
\caption{Test 2 for $\specstat_{Shannon}$, $\gamma=0.5$ and $\beta=0.8$. Plot shows $-\ln\E_{K}(\Delta_\specstat(\Xk^{(n)})^2)$
}
\label{fig:f_Shannon}
\end{figure}

For space considerations, the full documentation of our
results is deferred to the
Supporting Information \cite{SI}.
To offer a few examples,
Figure \ref{fig:KS1} and Table \ref{tab:KS1} show the linear fit to $\overline{\Delta}^{Var}_{KS}$ for  $(\gamma=0.5,\beta=0.8)$.
Figure \ref{fig:KS2} 
shows the linear fit to
$\overline{\Delta}^{Var}_{KS}$ for a different value of $\beta$, namely  $(\gamma=0.5,\beta=0.6)$.
Figure \ref{fig:f_AC} 
shows the linear fit to
$\overline{\Delta}_{\specstat_{AC}}$ for  $(\gamma=0.5,\beta=0.8)$.
Figure \ref{fig:f_Shannon} and Table \ref{tab:f_Shannon} show the linear fit to
$\overline{\Delta}_{\specstat_{Shannon}}$ for  $(\gamma=0.5,\beta=0.8)$. Similar figures and tables for the other values $(\gamma,\beta)$, in particular, $(\beta=0.3,\gamma=0.5)$, $(\beta=0.5,\gamma=0.5)$,  $(\beta=0.7,\gamma=0.5)$,  $(\beta=0.9,\gamma=0.5)$, $(\beta=0.6,\gamma=0.25)$, $(\beta=0.8,\gamma=0.25)$, are deferred to the Supporting Information. 
Note that in all coefficient tables, both those shown here and those deferred to the
Supporting Information, upper box shows complex frames (with t-test comparison
to the complex MANOVA ensemble of the same size,
denoted ``MANOVA'') and bottom box shows 
real frames (with t-test comparison to the real MANOVA ensemble of the same
size, denoted ``RealMANOVA'').
In each box, top rows are deterministic frames and bottom rows are random
frames.  Further note that in plots for Test 2 the horizontal axis is slightly
different for real and complex frames, as the preliminary step described above
was performed separately for real and complex frames. In the interest of space,
we plot all frames over the horizontal axis calculated for complex frames.

\paragraph{Validation on random frames.}
While our primary interest was in deterministic frames, we included in the
frames under study random frames. For the complex Haar frame and random Fourier
frame, convergence of the empirical CDF of the spectrum to the limiting
MANOVA$(\beta,\gamma)$ distribution has been proved in
\cite{Farrell,Edelman}. To our surprise, not only was our framework validated on
the four random frames under study, in the sense of asymptotic empirical spectral
distribution, but all universality hypotheses {\bf H1--H6} were accepted (not
rejected at the 0.001 significance level, with very few exceptions).

\paragraph{Test results on deterministic frames.}
A tabular summary of our results, per hypothesis and per frame under study, is included for convenience in the Supporting Information. 
Universality Hypotheses {\bf H1--H3} were accepted on all deterministic frames.
for {\bf H1--H2}, 
convergence of the empirical spectral distribution to the MANOVA$(\beta,\gamma)$
limit has been observed in all cases. For {\bf H3}, the linear fit in all cases
was excellent with $R^2>0.99$ without exception, confirming the power law 
in \eqref{KS_conv:eq} and the polynomial 
decrease of $\overline{\Delta^2}_{KS}$
with $n$.
Universality Hypothesis {\bf H4} was accepted (not rejected) for deterministic
equiangular tight
frames (ETFs) 
at the 0.001 significance level, with few exceptions (see Table \ref{tab:KS1} below, as well as full results and summary table in the Supporting Information); it was rejected for
deterministic non-ETFs. For $\gamma=0.25$, Hypothesis {\bf H4} has also been accepted for the Alltop frame, see Supporting Information.  
Universality Hypothesis {\bf H5} was accepted for all deterministic frames, with
excellent linear fits ($R^2>0.97$ without exception), confirming the power law
in  \eqref{func_conv:eq}.
Universality Hypothesis {\bf H6} was accepted (not rejected) at the 0.001
significance level (and even 0.05 with few exceptions) for all deterministic frames.
For the reader's convenience, 
Table \ref{ExpSummary} summarizes the universal
exponents for convergence of the entire
spectrum ({\bf H4}) and the universal exponents for convergence of 
the functionals under study ({\bf H6}), for $(\beta,\gamma)=(0.8,0.5)$.
The framework developed in this paper readily allows tabulation of these new
universal exponents for any value of $(\beta,\gamma)$.
We have observed that the universal exponents are slightly sensitive to the random seed. However, exact evaluation of this variability requires very significant computational resources and is beyond our present scope. Similarly, some sensitivity of the p-values to random seed has been observed.

\begin{table}[h]
\centering
\input{SpectrumKStest_empiricalVar_gamma0_5_beta1_25_minN1000_pnas3.tex}

\caption{Results of Test 1 for $\gamma=0.5$ and $\beta=0.8$.}
\label{tab:KS1}
\end{table}

\begin{table}[h]
\centering
\input{FuncShannon_Mse_gamma0_5_beta1_25_minN1000_pnas3.tex}
\caption{Results of Test 2 for $\specstat_{Shannon}$, $\gamma=0.5$ and $\beta=0.8$}
\label{tab:f_Shannon}
\end{table}

	\paragraph{Reproducibility advisory.} 
	All the figures and tables in this paper, including those in the
	Supporting Information, are fully reproducible from our
	raw results and code deposited in the Data and Code Supplement 
	\cite{SDR}.

\section*{Discussion}
\label{sec:discussion}

\subsection*{The hypotheses}

Our Universality Hypotheses may be surprising in several aspects:
Firstly, the frames examined were designed to minimize frame bounds and
worse-case pairwise correlations. Still it appears that they
perform well when the performance criterion is based on spectrum of the
typical selection of $k$ frame vectors. 
Secondly, under the Universality Hypotheses, all these deterministic frames
perform exactly as well as random frame designs such as the random Fourier
frame. Inasmuch as frames are continuous codes, we find deterministic codes
matching the performance of random codes. 
Finally, the Hypotheses suggest an extremely broad universality property: many
different ensembles of random matrices asymptotic exhibit the limiting
MANOVA spectrum.

All of the deterministic frames under study satisfy the Universality Hypotheses (with Hypothesis {\bf H4} satisfied only for ETFs). This should not give the impression that {\em any} deterministic frame satisfies these hypotheses! 
Firstly, because the empirical measures of an arbitrary sequence of frames rarely converge (thus violating Hypothesis H1). 
Secondly, even if they converge, a too-simplistic frame design often leads to concentration of the lower edge of the empirical spectrum near zero, resulting in a non-MANOVA spectrum and poor performance.
For example, if the frame is sparse, say, consisting of some $m$ columns of the $n$-by-$n$ identity matrix, then a fraction $(n-m)/n$ of the singular values of a typical submatrix are exactly zero. 

The frames under study are all ETFs or near-ETFs, all with favorable frame properties. 
To make this point, we have included in the Supporting Information \cite{SI}
study of a low-pass frame, in which the Fourier frequencies included in the frame are the lowest ones. This is in contrast with the clever choice of frequencies leading to the difference-set spectrum frame (DSS). Indeed the low-pass frame does not have appealing frame properties. It's quite obvious from the results in the SI, as well as results regarding the closely related random Vandermonde ensemble
\cite{VandermondeSpectrum}, that such frames do not satisfy any of the Universality Hypotheses {\bf H2--H6}.

We note that convergence rates of the form \eqref{KS_conv:eq} and
\eqref{func_conv:eq} are known for other classical random matrix ensembles 
\cite{Gotze1,Gotze2,chatterjee,meckes}.

We further note that Hypotheses {\bf H1--H4} do not imply Hypotheses {\bf H5--H6}.
Even if the empirical CDF converges in KS metric to the limiting
MANOVA$(\beta,\gamma)$ distribution, 
functionals which are not continuous in the KS metric do not necessarily
converge, and moreover no uniform rate of convergence is a-priori implied.

\subsection*{Our contributions}

This paper presents a novel, simple method for approximate computation (with
known and good approximation error) of spectral functionals of $k$-submatrix
ensemble for a variety of random and deterministic frames, using
\eqref{main_for:eq}.  Our results make it possible to tabulate these 
approximate values, creating a useful resource for scientists. 
As an example, we include Table \ref{TabulateAccuracy}. This is a lookup table
for the value of the functional $\Psi_{AC}$ on the difference-set spectrum
deterministic frame family (DSS), listing by values of $n$ and $k$ 
the asymptotic (approximate) value 
calculated analytically from the limiting $f^{MANOVA}_{\beta,\gamma}$
distribution, 
and the standard approximation error. 

To this end we developed a systematic empirical framework,
which allows validation of \eqref{main_for:eq} and discovery of the exponents
there.  Our work is fully reproducible, and our framework is available (along
with the rest of our results and code) in the Code and Data Supplement
\cite{SDR}.  In addition, our results provide overwhelming empirical evidence
for a number of phenomena, which were, to the best of knowledge, previously
unknown:

\begin{enumerate}
	\item {\bf The typical $k$-submatrix ensemble of deterministic frames is an
		object of interest.}
		While there is absolutely no randomness involved in the submatrix $X_K$
		of a deterministic frame (other than the choice of subset $K$), 
		the typical $k$-submatrix appears to be an ensemble in its own right,
		with properties so far attributed only to random matrix ensembles --
		including a universal, compactly-supported
		limiting spectral distribution and convergence of the maximal (resp.
		minimal) singular value to the upper (resp. lower) edges of the limiting
		distribution.

	\item {\bf MANOVA$(\beta,\gamma)$ as a universal limiting spectral distribution.}
		Wachter's MANOVA$(\beta,\gamma)$
		distribution is the limiting spectral distribution of $\lambda(G_K)$, as
		$k/m\to\beta$ and $m/n\to\gamma$, 
		for the typical $k$-submatrix ensemble of deterministic frames 
		(including difference-set, Grassmannian, real Paley, complex Paley,
		quadratic chirp, spiked and sines, and spikes and Hadamard).
		The same is true for real random frames - random cosine transform and
		random Haar.

	\item {\bf Convergence of the edge-spectrum.} 
		For all the deterministic frames above, as well as for the random frames 
		(random cosine, random Fourier, complex Haar, real Haar), the 
		maximal and minimal eigenvalues of the $k$-typical submatrix ensemble 
		converge to the support-edges of the MANOVA$(\beta,\gamma)$ limiting 
		distribution. The convergence follows a universal power-law rate.

	\item {\bf A definite power-law rate of convergence for the
			entire spectrum of the MANOVA$(n,m,k,\Fc)$ ensemble to its 
		MANOVA$(\beta,\gamma)$ limit,} with different exponents in the real and
		the complex cases.

	\item {\bf Universality of the power-law exponents for the entire
		spectrum.} The complex deterministic ETF frames 
		(difference-set, Grassmannian,
		complex Paley)  share the power-law exponents with the
		MANOVA$(n,\m,k,\C)$ ensemble. The same is true for
		the complex random frames (random Fourier and
		complex Haar).
        The complex tight non-equiangular Alltop frame, which can be constructed for various aspect ratios, also share the power-law exponents with the
		MANOVA$(n,\m,k,\C)$ ensemble for $\gamma<0.5$.
		The real deterministic ETF frame 
		(real Paley) 
        shares the exponent with the 
		MANOVA$(n,\m,k,\R)$. The same is true for real random frames 
		(random cosine and real Haar).
        All non-ETFs under study, with $\gamma=0.5$, share different power-law exponents (slower convergence).

	\item {\bf A definite power-law rate of convergence for
		functionals} including $\specstat_{StRIP}$, $\specstat_{AC}$ and 
	$\specstat_{Shannon}$.
    
	\item {\bf Universality of the power-law exponents for functionals.} 
	For practically all frames under study, both random and deterministic, 
	the power-law exponents for functionals agree with those of the 
	MANOVA$(n,\m,k,\R)$  (real frames) and 
	MANOVA$(n,\m,k,\C)$ (complex frames).



\end{enumerate}
\subsection*{Intercepts}

Our results showed a surprising categorization of the deterministic and random
frames under study, according to the constant $C$ in \eqref{KS_conv:eq}, or
equivalently, according to the intercept (vertical shift) in the linear regression 
on $\log(n)$. Figure \ref{fig:KS1} and Figure \ref{fig:KS2} 
clearly show that the regression lines, while having identical slopes (as
predicated by Hypothesis {\bf H3}), are grouped according to their intercepts
into the following 
seven categories:
	Complex MANOVA ensemble and complex Haar (Manova, HAAR);
	Real MANOVA ensemble and real Haar (RealManova, RealHAAR);
	Complex ETFs (DSS, GF, ComplexPF);
	Non-ETFs (SS,SH,Alltop);
	Real ETF (RealPF);
	Complex Random Fourier (RandDFT), and 
	Real Random Fourier (RandDCT).

	Interestingly, intercepts of all complex frames are larger (meaning that the
	linear coefficient $C$ in \eqref{KS_conv:eq} is smaller) than those of all
	real frames. Also, the less randomness exists in the frame, the higher the
	intercept: intercepts of deterministic ETFs are higher then those of random
	Fourier and random Cosine, which are in turn higher than those of Haar
frames and the MANOVA ensembles.

%
%




\subsection*{Related work} \label{sec:related}

Farrell \cite{Farrell} has
conjectured that the phenomenon of convergence of the spectrum of typical
$k$-submatrices to the limiting MANOVA distribution is indeed much broader and
extends beyond the partial Fourier frame he considered.
A related empirical study was conducted by Monajemi et al \cite{Monajemi}.
There, the
authors considered the so-called sparsity-undersampling phase transition in
compressed sensing. This asymptotic quantity poses a performance criterion for
frames that interacts with the typical $k$-submatrix $\Xk$ in a manner possibly
more complicated than the spectrum $\lambda(\Gk)$. The authors investigated
various deterministic frames, most of which are studied in this paper, and
brought empirical evidence that the phase transition for each of these
deterministic frames is identical to the phase transition of Gaussian frames.
Gurevich and Hadani \cite{Gurevich2009} proposed certain deterministic frame
construction and effectively proved that the empirical spectral distribution of
their typical $k$-submatrix converges to a semicircle, assuming
$k=\m^{1-\varepsilon}$, a scaling relation different than the one considered
here.
\cite{R21} and \cite{R22} also considered deterministic frame designs,
chirp sensing codes and binary linear codes, with a random sampling.
In their design the aspect ratios are large (e.g., in \cite{R21}
$m \sim k^2$ and $n \sim m^2$), so the spectrum converges to the
Mar\u cenko-Pastur distribution.
Tropp \cite{Tropp_cond} provided bounds for $\lambda_{max}(\Gk)$ and
$\lambda_{min}(\Gk)$ when $X$ is a general dictionary.
%
%
Collins \cite{Collins} has shown that the spectrum 
of a matrix model deriving
from random projections has the same eigenvalue distribution of the MANOVA
ensemble in finite
$n$. Wachter \cite{wachter} used a connection between the MANOVA ensemble and
submatrices of Haar matrices to derive the asymptotic spectral distribution
MANOVA$(\beta,\gamma)$.

%
%
%
%

\section*{Conclusions}

We have observed a surprising universality property for the $k$-submatrix
ensemble corresponding to various well-known deterministic frames, as well as to
well-known random frames. The MANOVA ensemble, and the MANOVA limiting
distribution, emerge as key objects in the study of frames, both random and
deterministic, in the context of sparse signals and erasure channels. 
We hope that our findings will invite rigorous mathematical study of these 
 fascinating phenomena.

	In any frame where our Universality Hypotheses hold
	(including all the frames under study here),
	Figure \ref{fig:limiting_F} correctly
describes the limiting values of $f_{RIP}$, $f_{AC}$ and $f_{Shannon}$ and shows
that codes based on deterministic frames (involving no randomness and allowing
fast implementations) are better, across performance measures,
than i.i.d random codes. 

The empirical framework we proposed in this paper may be easily applied to new
frame families $X^{(n)}$ and new functionals $\Psi$, extending our results further and
mapping the frontiers of the new universality property. 
In any frame family, and for any functional, where our Universality Hypotheses hold, we
have proposed a simple, effective method for calculating quantities of the form 
$\E_K \Psi\left( \lambda(G_K \right))$ to known approximation, which improves
	polynomially with $n$. 




	\subsection*{Acknowledgements} 
	The authors thank David Donoho for numerous helpful suggestions and the anonymous referees for their helpful comments. 
This work was partially supported by Israeli Science Foundation grant no.
1523/16.


\begin{sidewaystable}[htb]
   \centering
\input{ExponentsSummaryTable4.tex}
\caption{Summary of universal exponents for convergence. $\gamma = 0.5$, $\beta = 0.8$, $(\specstat_{S}=\specstat_{Shannon})$.
\label{ExpSummary}
}
\end{sidewaystable}

\begin{sidewaystable}[htb]
		\centering
\input{TabulateFlactuations_DSS_gamma0_5_3.tex}
\caption{$\Psi(f^{MANOVA}_{\beta,\gamma})\pm\sqrt{\Delta_\specstat(\Xkn^{(n)};n,\m_n,k_n)^2}$ for $\specstat_{AC}$ and DSS frame, $m=\frac{n-1}{2}$, $k=\beta\cdot\m$.
  }
		\label{TabulateAccuracy}
	\end{sidewaystable}

\end{document}

%% file: SpectrumKStest_empiricalVar_gamma0_5_beta1_25_minN1000_pnas3.tex
\begin{tabular}{|c|c|c|c|c|}\hline 
Frame & $R^2$& $\hat{b}$ & $SE(\hat{b})$ & p-value \\
 & & & & $b=b_{MANOVA}$\\ \hline 
MANOVA & 0.99828 & 0.92505 & 0.00690 & 1 \\ 
DSS & 0.99858 & 0.93652 & 0.00911 & 0.32089 \\ 
GF & 0.99921 & 0.92474 & 0.02608 & 0.99082  \\ 
ComplexPF & 0.99950 & 0.92454 & 0.00535 & 0.95390  \\ 
Alltop & 0.98906 & 0.49660 & 0.00883 & 9.4651e-47  \\ 
SS & 0.98767 & 0.47354 & 0.00950 & 5.8136e-45  \\ 
\hline 
HAAR & 0.99736 & 0.94421 & 0.00873 & 0.09019  \\ 
RandDFT & 0.99544 & 0.94127 & 0.01644 & 0.36788  \\ 
\hline 
\hline 
RealMANOVA & 0.99873 & 0.95610 & 0.00613 & 1 \\ 
RealPF & 0.99871 & 0.91244 & 0.00821 & 9.7174e-05  \\ 
SH & 0.99989 & 0.46822 & 0.00492 & 6.3109e-35  \\ 
\hline 
RealHAAR & 0.99596 & 0.94456 & 0.01081 & 0.35675  \\ 
RandDCT & 0.99773 & 0.93859 & 0.01156 & 0.18737  \\ 
\hline 
\end{tabular}

%% file: FuncShannon_Mse_gamma0_5_beta1_25_minN1000_pnas3.tex
\begin{tabular}{|c|c|c|c|c|}\hline 
Frame & $R^2$& $\hat{b}$ & $SE(\hat{b})$ & p-value \\
 & & & & $b=b_{MANOVA}$\\ \hline 
MANOVA & 0.98721 & 1.79936 & 0.03678 & 1 \\ 
DSS & 0.99110 & 1.88674 & 0.04615 & 0.14551 \\ 
GF & 0.99997 & 1.88548 & 0.01073 & 0.03161  \\ 
ComplexPF & 0.99977 & 1.77783 & 0.00701 & 0.56808  \\ 
Alltop & 0.93841 & 1.70618 & 0.07388 & 0.26297  \\ 
SS & 0.95539 & 1.89501 & 0.07355 & 0.24922  \\ 
\hline 
HAAR & 0.97971 & 1.87082 & 0.04836 & 0.24400  \\ 
RandDFT & 0.96928 & 1.77454 & 0.08157 & 0.78270  \\ 
\hline 
\hline 
RealMANOVA & 0.99202 & 2.05451 & 0.03309 & 1 \\ 
RealPF & 0.99834 & 2.00345 & 0.02045 & 0.19576  \\ 
SH & 0.97850 & 1.81297 & 0.26874 & 0.37904  \\ 
\hline 
RealHAAR & 0.98287 & 2.09078 & 0.04958 & 0.54503  \\ 
RandDCT & 0.98364 & 1.99663 & 0.06648 & 0.43977  \\ 
\hline 
\end{tabular}

%% file: ExponentsSummaryTable4.tex
\begin{tabular}{c|c|c|c|c|c|c|c|c|c|c|c|c|c} 
\toprule 
Frame & $b_{spectrum}$ & $b_{\specstat_{RIP}}$ &$a_{\specstat_{RIP}}$ &$b_{\specstat_{AC}}$ &$a_{\specstat_{AC}}$ &$b_{\specstat_{S}}$&$a_{\specstat_{S}}$ &$b_{\specstat_{max}}$&$a_{\specstat_{max}}$&$b_{\specstat_{min}}$&$a_{\specstat_{min}}$&$b_{\specstat_{cond}}$&$a_{\specstat_{cond}}$ \\
\hline ~
MANOVA & 0.93 & 1.15 & 2.21 & 1.44 & 3.48 & 1.80& 0.99 & 1.13 & 2.48& 1.00& 3.09 & 1.87& -4.55 \\ 
DSS & 0.94 & 1.14 & 2.18 & 1.40 & 3.40 & 1.89& 1.04 & 1.10 & 2.41& 1.00& 3.11 & 1.87& -4.56 \\ 
GF & 0.92 & 1.17 & 2.23 & 1.53 & 3.70 & 1.89& 1.03 & 1.13 & 2.48& 1.04 & 3.22 & 1.95& -4.76\\ 
ComplexPF & 0.92 & 1.13 & 2.17 & 1.44 & 3.49 & 1.78& 0.98 & 1.10 & 2.41& 1.00& 3.12 & 1.87& -4.56 \\ 
Alltop & 0.50 & 1.14 & 2.18 & 1.46 & 3.53 & 1.71& 0.94 & 1.11 & 2.42& 1.01 & 3.13 & 1.86& -4.54\\ 
SS & 0.47 & 1.11 & 2.13 & 1.50 & 3.63 & 1.90& 1.04 & 1.08 & 2.36& 0.98 & 3.06 & 1.83& -4.47\\ 
HAAR & 0.94 & 1.10 & 2.11 & 1.52 & 3.69 & 1.87& 1.03 & 1.09 & 2.37& 1.01& 3.13 & 1.88& -4.59 \\ 
RandDFT & 0.94 & 1.21 & 2.32 & 1.47 & 3.56 & 1.77& 0.97 & 1.11 & 2.42& 1.03& 3.18 & 1.93& -4.70 \\ 
\hline ~
RealMANOVA & 0.96 & 0.87 & 3.58 & 1.26 & 5.21 & 1.27& 5.26 & 0.90 & 3.73& 0.87 & 3.58 & 0.77& 3.17\\ 
RealPF & 0.91 & 0.92 & 3.82 & 1.32 & 5.46 & 1.24& 5.12 & 0.94 & 3.88& 0.94 & 3.88 & 0.81& 3.36\\ 
SH & 0.47 & 0.93 & 3.82 & 1.34 & 5.53 & 1.14& 4.71 & 0.93 & 3.82& 0.93& 3.82 & 0.85& 3.51 \\ 
RealHAAR & 0.94 & 0.86 & 3.54 & 1.23 & 5.07 & 1.29& 5.35 & 0.89 & 3.68& 0.90& 3.73 & 0.79& 3.28 \\ 
RandDCT & 0.94 & 0.99 & 4.08 & 1.30 & 5.38 & 1.24& 5.10 & 0.94 & 3.89& 0.95& 3.93 & 0.82& 3.40 \\ 
\bottomrule 
\end{tabular}

%% file: TabulateFlactuations_DSS_gamma0_5_3.tex
\begin{tabular}{|c|c|c|c|c|c|c|c|}\hline 
$n$ & 1031 & 1151 & 1291 & 1451 & 1571 & 1811 & 1951 \\ \hline 
RMSE, $\beta=0.8$ & 3$\pm$0.0281 & 3$\pm$0.0253 & 3$\pm$0.0227 & 3$\pm$0.0204 & 3$\pm$0.0189 & 3$\pm$0.0166 & 3$\pm$0.0155  \\ \hline 
RMSE, $\beta=0.6$ & 1.75$\pm$0.0073 & 1.75$\pm$0.0065 & 1.75$\pm$0.0058 & 1.75$\pm$0.0051 & 1.75$\pm$0.0048 & 1.75$\pm$0.0041 & 1.75$\pm$0.0038  \\ \hline 
\end{tabular}

%% file: arxiv_frames.bbl
\clearpage
\begin{thebibliography}{1}
        
\bibitem{forrester} Forrester, P. J. (2010).  \emph{Log-gases and random matrices (LMS-34)}. Princeton University Press.


\bibitem{Candes-RIP} Candes, E. J. (2008). The restricted isometry property and its implications for compressed sensing. \emph{Comptes Rendus Mathematique}, 346(9), 589-592.


\bibitem{AnalogCoding} Haikin, M., Zamir, R. (2016). Analog Coding of a Source with Erasures. In \emph{IEEE International Symposium on Information Theory Proceedings (ISIT)} (pp. 2074-2078). IEEE.


\bibitem{RandomMatrix} Tulino, A. M., Verd{\'u}, S. (2004). \emph{Random matrix theory and wireless communications} (Vol. 1). Now Publishers Inc.


\bibitem{Monajemi} Monajemi, H., Jafarpour, S., Gavish, M., Donoho, D. L. (2013). Deterministic matrices matching the compressed sensing phase transitions of Gaussian random matrices. \emph{Proceedings of the National Academy of Sciences}, 110(4), 1181-1186.


\bibitem{Farrell} Farrell, B. (2011). Limiting empirical singular value distribution of restrictions of discrete Fourier transform matrices. \emph{Journal of Fourier Analysis and Applications}, 17(4), 733-753.





\bibitem{WBdss}
Xia, P., Zhou, S., Giannakis, G. B. (2005). Achieving the Welch bound with difference sets. \emph{IEEE Transactions on Information Theory}, 51(5), 1900-1907.


\bibitem{Grassmannian}
Strohmer, T., Heath, R. W. (2003). Grassmannian frames with applications to coding and communication. \emph{Applied and computational harmonic analysis}, 14(3), 257-275.


\bibitem{PaleyConstruction}
Paley, R. E. (1933). On orthogonal matrices. \emph{Journal of Mathematics and Physics}, 12(1), 311-320.


\bibitem{Caulderbank-STRIP}
Calderbank, R., Howard, S., Jafarpour, S. (2010). Construction of a large class of deterministic sensing matrices that satisfy a statistical isometry property. \emph{IEEE journal of selected topics in signal processing}, 4(2), 358-374.





\bibitem{Johnstone2008}
Johnstone, I. M. (2008). Multivariate analysis and Jacobi ensembles: Largest eigenvalue, Tracy--Widom limits and rates of convergence. \emph{Annals of statistics}, 36(6), 2638.


\bibitem{Gurevich2009} Gurevich, S., Hadani, R. (2008). The statistical restricted isometry property and the Wigner semicircle distribution of incoherent dictionaries. \emph{arXiv preprint arXiv:0812.2602}.


\bibitem{MP} Mar{\v{c}}enko, V. A., Pastur, L. A. (1967). Distribution of eigenvalues for some sets of random matrices. \emph{Mathematics of the USSR-Sbornik}, 1(4), 457.


\bibitem{Oded} Haviv, I., Regev, O. (2016, January). The restricted isometry property of subsampled Fourier matrices. In \emph{Proceedings of the Twenty-Seventh Annual ACM-SIAM Symposium on Discrete Algorithms} (pp. 288-297). SIAM.


\bibitem{Donoho-stable} Donoho, D. L., Elad, M., Temlyakov, V. N. (2006). Stable recovery of sparse overcomplete representations in the presence of noise.  \emph{IEEE Transactions on information theory}, 52(1), 6-18.






\bibitem {Mazumdar} Mazumdar, A., Barg, A. (2011, July). General constructions of deterministic (s) rip matrices for compressive sampling. In \emph{IEEE International Symposium on Information Theory Proceedings (ISIT)} (pp. 678-682). IEEE.


\bibitem{Bandiera} Bandeira, A. S., Fickus, M., Mixon, D. G., Wong, P. (2013). The road to deterministic matrices with the restricted isometry property. \emph{Journal of Fourier Analysis and Applications}, 19(6), 1123-1149.


\bibitem{Fickus} Fickus, M., Jasper, J., Mixon, D. G., Peterson, J. (2015). Group-theoretic constructions of erasure-robust frames. \emph{Linear Algebra and its Applications}, 479, 131-154.


\bibitem{Rudelson} Rudelson, M., Vershynin, R. (2008). On sparse reconstruction from Fourier and Gaussian measurements. \emph{Communications on Pure and Applied Mathematics}, 61(8), 1025-1045.


\bibitem{CT06} Candes, E. J., Tao, T. (2006). Near-optimal signal recovery from random projections: Universal encoding strategies?. \emph{IEEE transactions on information theory}, 52(12), 5406-5425.


\bibitem{Nelson} Nelson, J., Price, E., Wootters, M. (2014, January). New constructions of RIP matrices with fast multiplication and fewer rows. In \emph{Proceedings of the Twenty-Fifth Annual ACM-SIAM Symposium on Discrete Algorithms} (pp. 1515-1528). SIAM.



\bibitem{Tropp} Pfander, G. E., Rauhut, H., Tropp, J. A. (2013). The restricted isometry property for time--frequency structured random matrices. \emph{Probability Theory and Related Fields}, 156(3-4), 707-737.


\bibitem{Edelman} Edelman, A., Sutton, B. D. (2008). The beta-Jacobi matrix model, the CS decomposition, and generalized singular value problems. \emph{Foundations of Computational Mathematics}, 8(2), 259-285.


\bibitem{Charaghchi} Cheraghchi, M., Guruswami, V., Velingker, A. (2013). Restricted isometry of Fourier matrices and list decodability of random linear codes. \emph{SIAM Journal on Computing}, 42(5), 1888-1914.


\bibitem{Collins} Collins, B. (2005). Product of random projections, Jacobi ensembles and universality problems arising from free probability. \emph{Probability theory and related fields}, 133(3), 315-344.




\bibitem{Lai} Foucart, S., Lai, M. J. (2009). Sparsest solutions of underdetermined linear systems via $\ell$q-minimization for 0$<$ q$\leqslant$ 1. \emph{Applied and Computational Harmonic Analysis}, 26(3), 395-407.


\bibitem{wachter} Wachter, K. W. (1980). The limiting empirical measure of multiple discriminant ratios. \emph{The Annals of Statistics}, 937-957.


\bibitem{silverstein_book}
Bai, Z., Silverstein, J. W. (2010). \emph{Spectral analysis of large dimensional random matrices} (Vol. 20). New York: Springer.


\bibitem {BaiBook} Yao, J., Bai, Z., Zheng, S. (2015). \emph{Large Sample Covariance Matrices and High-Dimensional Data Analysis} (No. 39). Cambridge University Press.


\bibitem {Tropp_cond} Tropp, J. A. (2008). On the conditioning of random subdictionaries. \emph{Applied and Computational Harmonic Analysis}, 25(1), 1-24.


\bibitem {Elad2010}
Elad, M. (2010). \emph{Sparse and Redundant Representations: From Theory to Applications in Signal and
Image Processing}. Springer New York.



\bibitem{Gotze1} G{\"o}tze, F., Tikhomirov, A. (2011). On the Rate of Convergence to the Marchenko--Pastur Distribution. \emph{arXiv preprint arXiv:1110.1284}.

    
\bibitem{Gotze2} G{\"o}tze, F., Tikhomirov, A. (2016). Optimal bounds for convergence of expected spectral distributions to the semi-circular law. \emph{aProbability Theory and Related Fields}, 165(1-2), 163-233.


\bibitem{chatterjee}
Chatterjee, S., Bose, A. (2004). A new method for bounding rates of convergence of empirical spectral distributions. \emph{Journal of Theoretical Probability}, 17(4), 1003-1019.

    
\bibitem{meckes} 
Meckes, E. S., Meckes, M. W. (2016). Rates of convergence for empirical spectral measures: a soft approach. \emph{arXiv preprint arXiv:1601.03720}.

\bibitem{VandermondeSpectrum} Debbah, M. (2008). Asymptotic Behaviour of Random Vandermonde Matrices with Entries on the Unit Circle. \emph{arXiv preprint arXiv:0802.3570}.

\bibitem{R21} Applebaum, L., Howard, S. D., Searle, S.,  Calderbank, R. (2009). Chirp sensing codes: Deterministic compressed sensing measurements for fast recovery. Appl. Comp. Harmonic Analysis, 26(2), 283-290.

\bibitem{R22} Babadi, B., Tarokh, V. (2011). Spectral distribution of random matrices from binary linear block codes. IEEE Trans. Info. Theory, 57(6), 3955-3962.

\bibitem{SDR}
	Code and Data Supplement for ``Random Subsets of Deterministic Frames have
	MANOVA Spectra''. Available online at \url{https://purl.stanford.edu/qg138qm8653}

\bibitem{SI}
	Supplamentary Information for ``Random Subsets of Deterministic Frames have
	MANOVA Spectra''. Available online at 
    \url{https://purl.stanford.edu/qg138qm8653}
    
\end{thebibliography}
